\documentclass[10pt,preprint2]{aastex}
\usepackage{bm}
\usepackage{lscape}
\usepackage[dvips]{epsfig}
\usepackage{amsmath}
\usepackage{amssymb}
\usepackage{amsthm}
\usepackage{array}
\usepackage{multirow}
\usepackage{graphicx}
\usepackage{amsfonts}
\usepackage{eucal}
\begin{document}

\shorttitle{}
\shortauthors{}


\newcommand{\ms}{$M_{\odot}$}
\newcommand{\msb}{$M_{\odot}$~}
\newcommand{\al}{$^{26}$Al}
\newcommand{\fe}{$^{60}$Fe}
\newcommand{\be}{$^{10}$Be}
\newcommand{\ca}{$^{41}$Ca}
\newcommand{\mn}{$^{53}$Mn}
\newcommand{\pd}{$^{107}$Pd}
\newcommand{\tc}{$^{99}$Tc}
\newcommand{\pu}{$^{244}$Pu}
\newcommand{\hf}{$^{182}$Hf}
\newcommand{\ct}{$^{13}$C}
\newcommand{\ctb}{$^{13}$C~}

\title{MHD and deep mixing in evolved stars.\\
I. 2D and 3D analytical models for the AGB}

\author{M.C. Nucci }
\affil{Department of Mathematics and Informatics, University of Perugia,  via Vanvitelli 1, Perugia, Italy, 06123; and
INFN, Section of Perugia, via Pascoli, Perugia, Italy, 06123; mariaclara.nucci@unipg.it}

\author{M. Busso}
\affil{Department of Physics and Geology, University of Perugia, and INFN, Section of Perugia,
via Pascoli, Perugia, Italy, 06123; busso@fisica.unipg.it}

\begin{abstract}
The advection of thermonuclear ashes by magnetized domains emerging from near the H-shell was suggested to explain AGB star abundances. Here we verify this idea quantitatively through exact MHD models. Starting with a simple 2D geometry and in an inertia frame, we study plasma equilibria avoiding the complications of numerical simulations. We show that, below the convective envelope of an AGB star, variable magnetic fields induce a natural expansion, permitted by the almost ideal MHD conditions, in which the radial velocity grows as the second power of the radius. We then study the convective envelope, where the complexity of macro-turbulence allows only for a schematic analytical treatment. Here the radial velocity depends on the square root of the radius. We then verify the robustness of our results with 3D calculations for the velocity, showing that, for both the studied regions, the solution previously found can be seen as a planar section of a more complex behavior, in which anyway the average radial velocity retains the same dependency on radius found in 2D. As a final check, we compare our results to approximate descriptions of buoyant magnetic structures. For realistic boundary conditions the envelope crossing times are sufficient to disperse in the huge convective zone any material transported, suggesting magnetic advection as a promising mechanism for deep mixing. The mixing velocities are smaller than for convection, but larger than for diffusion and adequate to extra-mixing in red giants.

\end{abstract}

\section{Introduction}

Stars below $\sim$ 2.2 \ms, while ascending the Red Giant Branch (RGB), display
an isotopic mix of elements up to oxygen that cannot be accounted for by
convective transport in the so-called {\it first dredge-up} \citep{gil}.
The same occurs in the subsequent Asymptotic Giant Branch (AGB) phases,
where the abundances do not follow model expectations for the {\it third dredge-up}
\citep{bus10}. Information on the above issues started to accumulate several years ago
\citep{har,gb,pil,kra}. Measurements on presolar grains of AGB origin confirmed
this evi\-dence, also show\-ing  high $^{26}$Al/$^{27}$Al ratios, inferred from
$^{26}$Mg excesses \citep{choi,amari,zin,nol}.

Radial transport of H-burning ashes is usually assumed to account for the
observational data \citep[see e.g.][and references
therein]{was,bus10,palm1,palm2}. Various physical processes  were explored for
such a transport, from rotational shear to gravitational waves \citep{zahn,dt03}.
Summaries can be found in \citet{pin97,dt03,d+06}.  Whatever causes mixing is expected
to explain also the evolution of $^3$He in the Universe \citep[see e.g. ][]{was,ban}
and, during the Main Sequence, the angular momentum transport accounting for
the slow rotational rate of the solar core \citep{tom},

Mixing phenomena induced directly by rotation were shown to be
ineffective \citep[see e.g.][]{pala,chala}, at least along the RGB. \citet{egg1,egg2} then pointed out that an alternative might
be offered by a diffusive phenomenon induced by $^3$He burning into $^4$He
and two protons, which decreases the average molecular weight. This leads to
a local molecular-weight inversion, an unstable situation in which
chemical and thermal readjustments are to be expected; the nature of the
ensuing mixing process was identified  by
\citet{cz07} with the known double-diffusive mechanism called
{\it thermohaline instability}. It was however later suggested \citep{dm11} that
the transport of chemicals thus induced is not efficient enough to account for
the observations. This is so especially for the most evolved AGB stars,
where presolar grains are formed \citep{palm1,palm2}.

Another possibility was identified by \citet{bus07} [hereafter BWNC] in the
advection of H-burning ashes through the buoyancy of magnetic flux tubes, descending
from a dynamo process. This follows old ideas by \citet{park84} for
the Sun \citep[see also][]{sp99,sp02}. Magnetized structures, even those that
are unstable in the lowest envelope layers of the Sun \citep{sb82a,sb82b,mi92},
might actually dissipate into the huge convective envelopes of evolved stars
\citep{hs01}, thus explaining also the absence of AGB coronae. In fact, the treatment
by BWNC would remain essentially valid for any form of magnetized domains
\citep[including e.g. the instabilities developing into $\Omega$-shaped loops extensively discussed by][]
{park74,park77,park84,park94}. Recent upgrades of this scenario can be
found in \citet{web}.

If the magnetized material starts its upward motion in regions close to the H-burning shell
and is characterized by the chemical peculiarities generated there, it can induce abundance
anomalies at the surface. The density unbalance produced by the upflows would
also generate downflows for maintaining mass conservation \citep{park84}, thus providing
the required circulation. Recently, on the basis of numerical
$\alpha-\Omega$ dynamo models, \citet{n+08} showed that,
with only a small energy drain from the convective envelope, the differential
rotation required to maintain a dynamo and the magnetic buoyancy
can be sustained. Using an appropriate velocity field, these
authors also showed that magnetic field values similar to those envisaged by
BWNC at the base of the convective envelope would result from the
dynamo process. Subsequently, attempts have been done
to merge the ideas of thermohaline and magnetic mixing, suggesting
magneto-thermohaline mechanisms \citep{d+9,dm11}.

Modelling in three dimensions the dynamo process from which magnetically-induced matter transport originally derives
\citep[see e.g.][]{noza,n+08,pas} is a task nor\-mally under\-taken with strong mathematical
simplifications, often introduced through a first-order perturbation treatment. In a problem
that is intrinsically highly nonlinear, such approaches,  unavoidable in complex 3D
calculations, are highly uncertain. Due to the possible relevance of magnetic mechanisms for
stellar mixing, this paper aims at verifying whether the advection of matter by magnetic fields
is really a viable transport mechanism, i.e. if it offers
an exact solution to the MHD equations. In so doing, we start with a simple 2D geometry,
studying the motion using polar coordinates in planes parallel to the
equator, as finding complete and {\it exact} 3D solutions in an analytic way is a very complex 
mathematical task. It is also hardly constrained physically, because one should describe the 
interactions between poloidal and toroidal fields, something for which we do not know anything experimental or observational in AGB stars.  On the other hand, trusting 2D solutions is risky if not properly verified with a 3D scheme. As a compromise between two opposite difficulties, we shall therefore perform a dedicated 3D analytical approach to the problem that is, for us, most important: i.e. the behavior of the radial velocity. In this way, we aim at verifying under which conditions, if any, the solution found in the 2D framework continues to hold in 3D, thus confirming or not magnetic buoyancy as an effective promoter of radial matter 
transport.

The need for non-convective transport affects other problems of stellar physics, beyond those to which this work is dedicated. One such case, closely connected to the phases discussed here, is the penetration of protons from the envelope into the He-rich layers when the H-burning shell is switched off, after a thermal pulse. In that case the transport is supposed to drive the subsequent formation of $^{13}$C in a local reservoir, where it will then burn through the $^{13}$C($\alpha$,n)$^{16}$O reaction, producing the neutrons needed for s-processing. We consider therefore the present attempt as being also preliminary for subsequently dealing with that second problem (which will be presented separately in a forthcoming paper, as the part number II of our entire work; it cannot actually be included  here for a question of space).

In Section 2 the MHD equations are introduced and our assumptions for the geometry of the magnetic field and for the environmental conditions are presented. The exact analytical solutions for the equilibrium of a magnetized stellar plasma are shown in subsection 2.1 (their derivation is also synthetically outlined in the Appendix). Then subsection 2.2 illustrates how these solutions naturally imply an expansion if the magnetic field is variable in time and how this relates to the usual treatment of buoyancy in stellar physics. In Section 3 the specific case of the radiative layers above the H-burning shell of an evolved star is discussed in some detail.  Section 4 then presents a general discussion of what is expected to occur in the envelope. Here the presence of macro-turbulence related to convection, which cannot be treated exactly, permits only a rather schematic and time-independent approach. In section 5 we extend the analysis considering also meridional motions, i.e we pursue a 3D modelling, to understand under which conditions the solution found for the radial velocity can
continue to hold. Finally, some general implications of our analysis and their encouraging indications in favor of magnetically-induced mixing are summarized in Section 6.

\section{MHD with azimuthal fields in red giants}

The equations of the problem, expressed in Eulerian form and adopting cgs units are:
$$
\frac{\partial \rho}{\partial t} + \nabla \cdot(\rho
{\mathbf v}) = 0 \eqno (1)
$$
$$
{\rho \Big[\frac{\partial{\mathbf v}}
{\partial t} +
({\mathbf v}\cdot\nabla){\mathbf v} - c_d{\mathbf v} + \nabla \Psi\Big] - \mu \Delta{\mathbf v}} + \nabla P
$$
$$
+ \frac{1}{4\pi}{\mathbf B}\times(\nabla\times{\mathbf B})
= 0 \eqno (2)
$$
$$
\frac{\partial {\mathbf B}}{\partial t} - \nabla \times
({\mathbf v} \times {\mathbf B}) - \nu_m \Delta {\mathbf B} = 0 \eqno (3)
$$
$$
\nabla \cdot {\mathbf B} = 0     \eqno (4)
$$
$$
\rho \Big[\frac{\partial \epsilon}{\partial t}+
({\mathbf v}\cdot\nabla) \epsilon\Big] + P \nabla
\cdot {\mathbf v} - \nabla \cdot (\kappa \nabla T)
$$
$$
+ \frac{\nu_m}{4 \pi}(\nabla\times {\mathbf B})^2 = 0 \eqno (5)
$$
\setlength{\textwidth}{6.2in} \setlength{\textheight}{8in}
In the above equations, $\epsilon$ is the internal energy per unit mass. $P, T, \rho$  are the
pressure, temperature and density of the plasma, $\kappa$
is the thermal conductivity. ${\mathbf B}$ is the magnetic induction
field, ${\mathbf v}$ is the plasma velocity, $\mu$ is the
$dynamic ~viscosity$ (product of density and of the kinematic viscosity $\eta$) and
$\mu \Delta {\bf v}$ is a simplified form often used for the viscous force
per unit volume in stellar MHD (it would formally hold for incompressible fluids with constant $\mu$). $\Psi$ is the gravitational potential and $\nu_m$ is the $magnetic ~diffusivity$. The term $c_d {\bf v}$ represents the aerodynamic drag force per unit mass.

\vspace{0.7cm}
\begin{table}[t!!]
\small
\begin{center}
\begin{tabular}{|l|r|}
\multicolumn{2}{c}{TABLE I}\\
\multicolumn{2}{c}{Parameters of the AGB Star Layers of Interest}\\
\multicolumn{2}{c}{M=1.5 M$_{\odot}$, Z=0.01}\\
  \hline
Parameter & Value \\
  \hline
  $r_P$  & 1.97$\cdot$10$^9$ \\
  $\rho_P$ & 4.13 \\
  $T_P$ & 4.92$\cdot 10^7$ \\
  $P_P$ & 4.24$\cdot 10^{16}$ \\
  $r_{env}$ & 5.19$\cdot 10^{10}$ \\
  $\rho_{env}$ & $2.48\cdot 10^{-4}$ \\
  $k_{rad}$ & $-3$ \\
  $T_{env}$ & 2.17$\cdot 10^6$ \\
  $P_{env}$ & 1.29$\cdot 10^{11}$ \\
  $k_{con}$ & $-3/2$ \\
  $r_{sur}$ & 2.32$\cdot 10^{13}$ \\
  $T_{sur}$ & 3$\cdot 10^3$ \\
  $P_{sur}$ & $\approx 5 \cdot 10 ^{-2}$ \\
  $\rho_{sur}$ & $\approx$ 10$^{-9}$\\

  \hline
\end{tabular}
\end{center}
\vspace{-2mm} \linespread{0.8} \small{Notes: units are cgs. Concerning the subscripts, ``P"
refers to the maximum mixing penetration; ``env" refers to the border between the
radiative layer and the convective envelope; ``sur" indicates the surface values. The
physical parameters $T$, $\rho$, $P$, $r$ are from BWNC.}
\end{table}

\subsection{The equilibrium of a stellar plasma in the quasi-ideal MHD case}
In the light of the approach outlined in the Introduction,
we start our analysis using a simple 2D geometry for the fields, in an inertia frame, 
hence avoiding effects
like poleward tilts induced by the apparent Coriolis force \citep[see e.g.][]{cg87}.
Indicating with $r$ the radial coordinate and with $\varphi$ the azimuthal angle
in the equatorial plane, we therefore assume that ${\mathbf
B}=(B_r(t,r,\varphi),B_{\varphi}(t,r,\varphi),0)$ is such that $B_r$ = 0. This
describes an azimuthal field as a function of $r$, $\varphi$ and time $t$. We
also consider pure circular symmetry in the equatorial plane, so that the
velocity components do not depend on the azimuthal angle; let also the velocity field be
parallel to the equator. Hence: ${\mathbf
v}=(v_r(t,r,\varphi),v_{\varphi}(t,r,\varphi),0)$ is such that $v_{\varphi} =
v_{\varphi}(t,r)$ and $v_{r} =  v_{r}(t,r)$.

Let us first discuss the conditions under which we want to consider the induction
equation (3). In several astrophysical dynamo scenarios, the third term is much
smaller than the second one, as the conductivity of a ionized medium is very large.
Magnetic diffusivity is actually negligible if we consider transport phenomena occurring
through advection, by the fields described in the second term of the equation
\citep[it is common to say that, in this situation, the field and the transported fluid are
mutually ``frozen", and the magnetic flux through any co-moving closed circuit remains
constant: see e.g. Chapter 9 in][]{sch}. 

As examples, we can remember that for
the Earth the second term of equation (3) has the typical time scale of 60 yr, the third
one of 10$^4$ yr; for the Sun the second term is of the order of weeks, the third one
is 22 yr. In order to explain relatively fast mixing episodes in stars, avoiding the
problems previously encountered with the slow  thermohaline diffusion, we need here to
address the peculiar property of magnetic mechanisms of making in principle possible
relatively fast advection, in which matter and fields are almost frozen with respect
to each other (i.e. we have a case of almost ideal MHD).

In the layers above a H-burning shell of an evolved red giant (which are rather peculiar as compared to normal radiative regions in stars), the above situation actually occurs. In particular, the thermal and magnetic diffusivities are similar and both are much smaller than the (kinematic) viscosity. This is expressed by the fact that the so-called {\it magnetic Prandtl number} $P_m = \eta/\nu_m$ is much larger than unity. Its actual value can be derived with reference to a model star, for example that discussed by BWNC, whose basic characteristics are reported in Table 1. Using the expression $P_m \simeq 2.6\times 10^{-5}T^4/n$, where $n$ is the number density \citep{spi,sce}, it is easy to show that in the deepest layer possibly interested by non-convective transport \citep{nol}, $P_m$ is of the order of 60 (this layer is identified in Table 1 by the label ``P").  Notice that the thermal conductivity and the magnetic diffusivity act in a similar way, through second-order derivatives, namely $\Delta T$, $\Delta B$, respectively. In our hypotheses we assume therefore that thermal exchanges and magnetic diffusion are negligible ($\nu_m \simeq $ 0 and $\kappa \simeq $ 0) over the radial transport time scale. This also requires that the dynamic viscosity must be small, otherwise the viscous term would produce significant heating; but since $P_m$ indicates that $\eta$ is more important than $\nu_m$, we cannot formally neglect the viscous term in our equations. We have therefore a ``quasi-frozen" case. We shall then verify ({\it a posteriori}) if the results we find are consistent with the initial hypotheses, i.e. if the resulting motions do really operate on short enough time scales that the ``almost-frozen" field assumption is justified (see the end of Section 6 for this).

In the absence of macroturbulence, an approximate expression for $\mu$ and $\eta$ can be obtained from the particle scattering processes. Starting with neutral particles, let $c_s$ be the sound speed and $\lambda \simeq 1/(\sigma n)$  be the mean free path, where $n$ is the number density and $\sigma$ is the cross section; then: $\mu \simeq \rho \lambda c_s$. Adopting for $\sigma$ the value $\pi r_B^2$, where $r_B$ is the DeBroglie wavelength for the proton, we can estimate $\mu$ by computing $r_B$ and $c_s$ from the parameters of Table 1 (at the layer ``P"). It is easy to get $\mu \simeq 0.01$.  In most cases of interest for ionized plasmas, the Coulomb interaction among charged particles is also crucial; this is true within a distance from an ion shorter than a few Debye's radii (beyond this distance electron screening makes the plasma appear as essentially neutral). A quick calculation shows however that, in the stellar layers we consider, Coulomb interactions are actually not very important for viscosity. The Debye radius $\lambda_D$ for these zones was computed recently by \citet{sim}. From their Table 2 one gets that at the level ``P" $\lambda_D$ is of 1$-$2 atomic units ($\sim 10^{-8}$cm). On the other hand, the mean free path of a particle against collisions is
about 10$^{-5}$ cm, i.e. a value three orders of magnitude larger than the Debye radius. Hence for the largest part of their trajectory before undergoing a Coulomb interaction, particles will behave as they were essentially neutral (this is one of the consequences
of the so-called {\it screening} effects in stellar plasmas). We therefore remain with the small viscosity value computed above
($\mu \simeq 0.01$). We shall use this later for verifying that our assumptions of quasi-ideal MHD are realistic (Section 6).

Obviously, the above conditions (both for the viscosity and for the other dissipative terms) are far from general. For example, they are certainly not valid (by definition) for any kind of transport in which diffusive (slow) processes are important. They are also not true for convective regions, where the existence of macro-turbulence can induce processes of instability and disruption of the magnetized structures on time scales depending on the length scale of the convective eddies
\citep[see e.g. the discussion by][and our comments on the envelope situation in Section 5]{vish}.
For our purposes it is however important that they hold in the specific layers and evolutionary
phases where deep mixing is needed in stars (it is easy to show that they also hold in the same
layers of red giants after the so-called {\it bump} of the Luminosity Function).

In stellar applications the MHD equations are coupled to those
describing the physics of the star; the resulting system of nonlinear partial
differential equations is then very complex. The behavior of magnetized zones
is usually attacked in some simplified way, dealing
with the dynamics of the system after assuming the so-called {\it thin flux tube}
approximation \citep{sp81}. This last consists in making an expansion of MHD in
powers of $a/D$, where $D$ is a characteristic length scale for the variation
of the parameters and $a$ is a typical size of the most remarkable field structures forming,
looking like flux tubes in fluid dynamics. The parameter $a$ can be actually
seen as being the diameter of such magnetized tubes.  The expansion is
then stopped at the lowest order. In most approaches, the effects introduced by
magnetic fields on the general pressure stratification are assumed to be small
and the dynamic equations for the tubes are linearized \citep{sb82a,sb82b}.
The reference system usually adopted is at rest with any non-expanding ({\it neutral}) gas and  rotates with it. This non-inertiality requires consideration of {\it apparent} forces, like the
Coriolis force, perpendicular to gravity, so that also in treating the density this requires
considering lateral or azimuthal components. For a discussion
of this approach see e.g. \citet{fan06}; in particular Figure 5 of that paper illustrates well  the force balance.  \cite[See also][and references therein]{remp, park74, park84, d+9}. A general review specifically devised for the solar conditions can be found in \citet{sol}.

We have already mentioned that we aim at presenting here a different, complementary view.
In particular, our approach is similar to that of the seminal papers \citep[][]{park58,park60}
where the solar wind was shown to derive from the natural dynamic equilibrium
of a hot corona (the main differences being our explicit inclusion of magnetic effects
and the use of two dimensions instead of a purely spherical geometry). In practice, we don't
apply the usual simplifications in solving the MHD equations (hence we avoid the thin flux tube
approximation, which may induce further uncertainties). Instead, we pursue an exact analytical
approach, but in a simplified geometry, obtained: i) by excluding rotation (as in the quoted
Parker's work); and ii) by looking for 2D solutions, verifying then the results with a dedicated 
3D extension.
\begin{figure}[t!!]
\centerline{\includegraphics[width=\columnwidth]{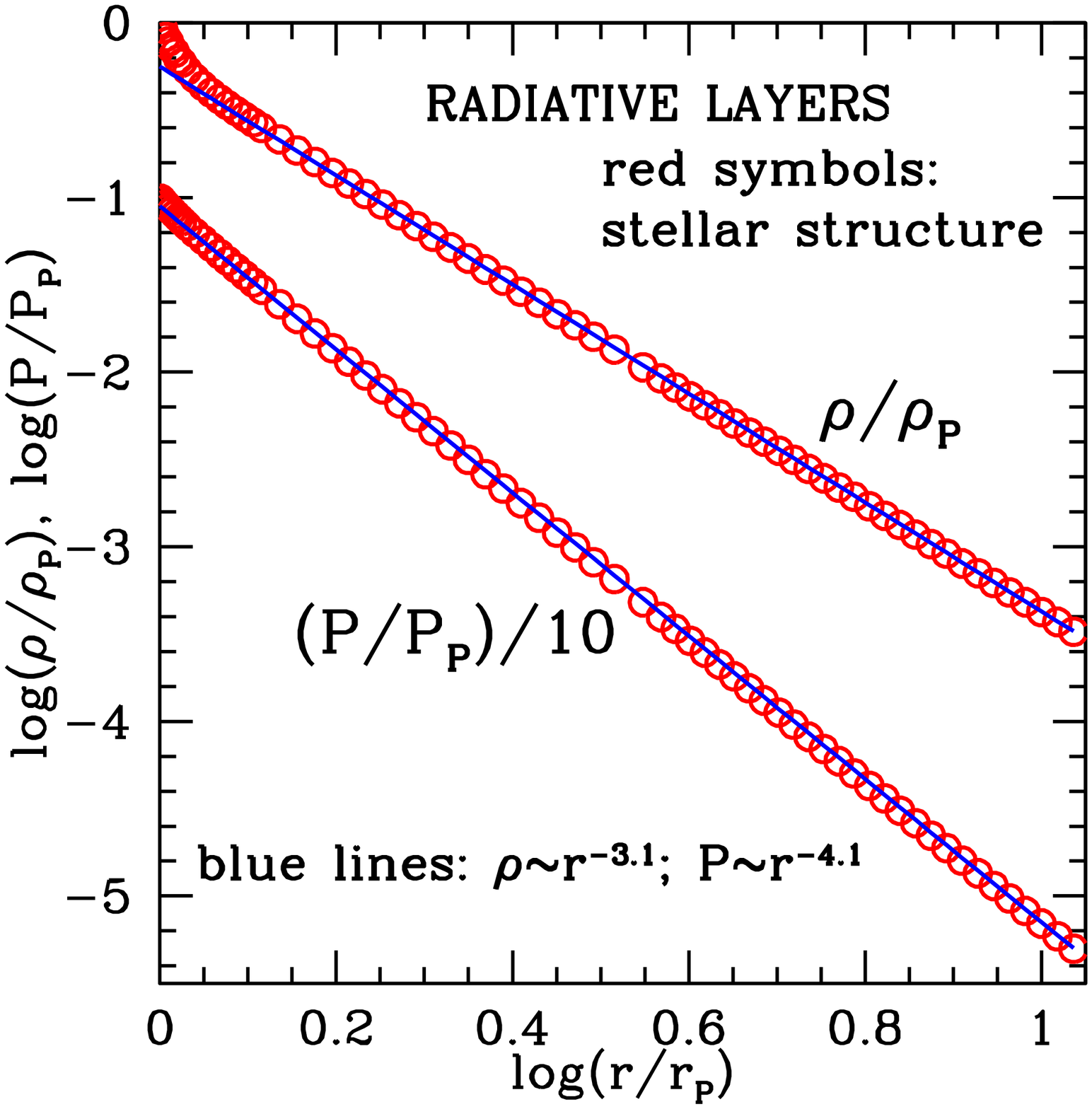}}
\vspace{1.5cm}
\caption{The dependency of density and pressure on radius in the radiative layers above the H-burning shell, in a 1.5 \msb star during the AGB phases (in the third interpulse period), according to the stellar models by \citet{str03}, which were adopted by BWNC and in recent discussions of deep mixing in evolved stars by \citet{palm1,palm2}. Note how the best fits are very close to a polytropic structure of index 3, as $\rho \sim (r/r_P)^{-3}$ and $P \sim (r/r_P)^{-4}$, so that $P \sim \rho^{4/3}$.}
\label{prhorad}
\end{figure}

The above choices prevent us to derive general conclusions on azimuthal motions and on meridional
circulation. This is for us acceptable, as our goal in only to understand under which conditions (if any)
the equilibrium of a magnetized stellar plasma, derived exactly, predicts a radial expansion, both in 2D and in 3D.
We aim further at finding the radial velocity profile that is to be expected, independently of any assumption
on the field organization and geometry. Only later (Section 6) we shall discuss the implications of our
solutions for practical cases, by fixing the parameters at values suggested either by
stellar physics or by previous studies of magnetic fields in red giants, including the
thin flux tube approximation.

\begin{figure}[t!!]
\centerline{\includegraphics[width=\columnwidth]{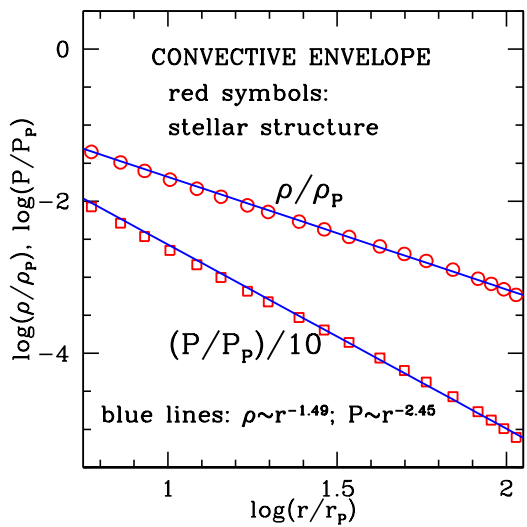}}
\vspace{1.5cm}
\caption{The dependency of density and pressure on radius in the central parts of the convective envelope, for the same 1.5 \msb AGB star of Figure 1. Note how the best fits are very close to a polytropic structure of index 1.5, as $\rho \sim (r/r_{env})^{-1.5}$ and $P \sim (r/r_{env})^{-2.5}$, so that $P \sim \rho^{5/3}$).}
\label{prhocon}
\end{figure}

In the above procedure, for the stellar structure we shall refer to the AGB model studied by
BWNC, whose parameters are summarized in Table 1. That model shows how the density distribution as
a function of the radius is very close to a power law, $\rho = \rho_0 (r/r_0)^{k}$ with $k \simeq -$3
in the radiative layers (see Figure \ref{prhorad}) and $k \simeq -$3/2 in the bulk of the convective
envelope (see Figure \ref{prhocon}).

In such conditions the solution of the MHD equations (1), (2), (3) and (4) describes the equilibrium of the stellar plasma in an inertia frame.
A brief outline of how this solution is obtained is presented in the Appendix.

The {\it mathematical} solution described in the Appendix is the following:
$$
v_r=\displaystyle\frac{{\rm d}w(t)}{{\rm d}t}\, {r}^{-(k+1)} \eqno(6)
$$
$$
B_{\varphi}=\Phi(\xi){r}^{k+1},\quad \left[\xi=-(k+2)w(t)+{r}^{k+2}\right]
\eqno(7)
$$
Here, $w(t)$ and $\Phi(\xi)$ are functions of $t$ and $\xi$, which are mathematically arbitrary, but require physically to be specified (in such a way that they also maintain the proper dimensions for $v_r$ and $B_{\varphi}$).

By avoiding any assumption on the field organization these solutions provide values for the average fields in the plasma $<B_{av}>$. 

In practical stellar situations, the ratio of the gas
pressure to the magnetic pressure is much higher than unity; this means that $<P>$/($<B^2>/8\pi$)=$\beta>>$1. As mentioned, in such conditions most works describe the field organization in flux tubes. In order to compare our results to those of numerical simulations adopting this approach one can remember that the average field in flux tubes $<B_t>$ is related to the average field in the plasma $<B_{av}>$, at pressure equilibrium, by the relation:
$$
<B_{av}^2> \sim \frac{1}{\beta} <B_t^2> \eqno(8)
$$
as shown by \citet{n+08}.

\subsection{The dynamical plasma equilibrium}
 Our solution (relations 6 and 7), applies to any plasma satisfying our boundary conditions;
 for it, the momentum equation (2), i.e. the classical Navier-Stokes equation, includes
 all mechanical and electromagnetic forces, as appearing in an inertia frame (we hence exclude the apparent forces commonly induced
 in studying a {\it relative} motion in a rotating frame). Magnetic effects are accounted for by the last term, which includes a double
 vectorial product. It can be split mathematically in two parts \citep[see e.g.][]{sch}. The first one is proportional to  $({\bf B} \cdot \nabla){\bf B}$ and describes a magnetic tension along the field lines. The second term is proportional to $\nabla B^2$. As $B^2/(8 \pi)$ is the magnetic
 pressure, this refers to the magnetic pressure gradient. It is common, in stellar physics, to provide illustrative examples (see e.g. BWNC and references therein) that reduce the enormous complexity of the magnetic field organization to a scheme  including a {\it magnetized} phase and a {\it non-magnetized} one. In such a way, the plasma inside a magnetized region (indicated by the subscript $``i"$) will experience both a gas and a magnetic pressure. At pressure
 equilibrium, their sum must equal the pressure of the external gas (indicated by the subscript $``e"$), in which the field is not present. Hence $\rho_i T_i < \rho_e T_e$ and if temperature equilibrium is achieved, then $\rho_i < \rho_e$, i.e. the magnetized domain moves toward regions of lower density. As a consequence, one says that magnetic fields induce a magnetic buoyancy force per unit length proportional to $g (\rho_e - \rho_i)$. By using our complementary approach in an intertia frame, we can emphasize properties of the plasma that are not otherwise obvious.
 This can be shown  by considering our velocity field (6), which is connected to the magnetic field through the function $w(t)$. Suppose now that $w(t)$ = constant. In this case its derivative is zero and $v_{r}(r,t)$ = 0. A plasma with a constant magnetic field (not necessarily with {\it zero} magnetic field) allows therefore for a static solution. Another interesting property is revealed by chosing $\Phi(\xi)$ = const: in this case $w(t)$ is arbitrary, thus allowing also for oscillating behaviors (see next Section). A static equilibrium is in any case not possible if $B$ varies in time: hence our treament makes clear that a plasma with a variable $B(t)$ has naturally a {\it dynamic} equilibrium and is in expansion. This is similar to the case of an expanding stellar corona, as studied in the already quoted classical papers \citep{park58,park60}.

 In our case expansion is secured unless $k > -1$ (in which situation the velocity
 drops with radius): this case is however never met in hydrostatic stellar evolution.
 We notice that the magnetically-induced expansion adds to the intrinsic stability
 problems of the peculiar radiative layers of an AGB star, whose thermodynamic structure
 (a polytrope of index 3, like for a bubble of radiation) approaches the Eddington's
 instability limit. This point deserves further scrutiny,
 as there is a possibility that the extra momentum induced by magnetic fields becomes an important ingredient for treating the
 still elusive problem of the envelope ejection in the pre-planetary nebula stage.

Summing up, the exact analytical approach proposed in Section 2.1 and in the Appendix, although performed in a simple geometry, accounts
naturally for an expansion of the magnetized plasma below the envelope of an AGB star. It
also makes clear that this must occur if the fields vary in time: this is actually a sufficient condition.
Working in an inertia frame also allows us to study the above equilibrium without reference to idealized non-physical concepts (like the density $\rho_e$ of a non-magnetized stellar gas) and
without the need of dealing explicitly with rotation and with the apparent forces induced by it.
We notice that, for zero magnetic field, cases of pure hydrodynamical expansion are also possible, if the derivative of $w(t)$ is non-zero.
Hence our description is more general than immediately evident and other specific dynamical
properties are a priori included: they will obviously depend on the assumed boundary conditions. 

\section{Magnetically-driven expansion in the radiative layers}

We recall that in the layers below the convective envelope, where extended mixing must occur,
any coupling of nucleosynthesis and circulation is constrained by the requirement
that the energy budget of the star is not modified, which fact implies the
empiric rule that mixing does not penetrate layers deeper than those with
$Log~T_H - Log~T_P \ge 0.1$, where $T_H$ is the temperature of the H-burning
shell and $T_P$ is the temperature at the maximum level of penetration
\citep{nol}. We call $\rho_P$, $r_P$ etc the values of the
variables and functions in this layer of maximum penetration (Table 1).
As already mentioned, we also have:
$$
\varrho=c_0 r^k \eqno (9)
$$
where $c_0 = \varrho_P/r_P^k$ and $k = -3$ (see Figure 1). 

The solution presented in the
previous Section now becomes:
$$
v_r=\displaystyle\frac{{\rm d}w(t)}{{\rm d}t}\, {r}^2 \eqno(10)
$$
$$
B_{\varphi}=\Phi(\xi){r }^{-2},\quad \left[\xi = w(t)+{r}^{-1}\right] \eqno(11)
$$

\begin{figure*}[t!!]
\centerline{\includegraphics[width=0.9\textwidth]{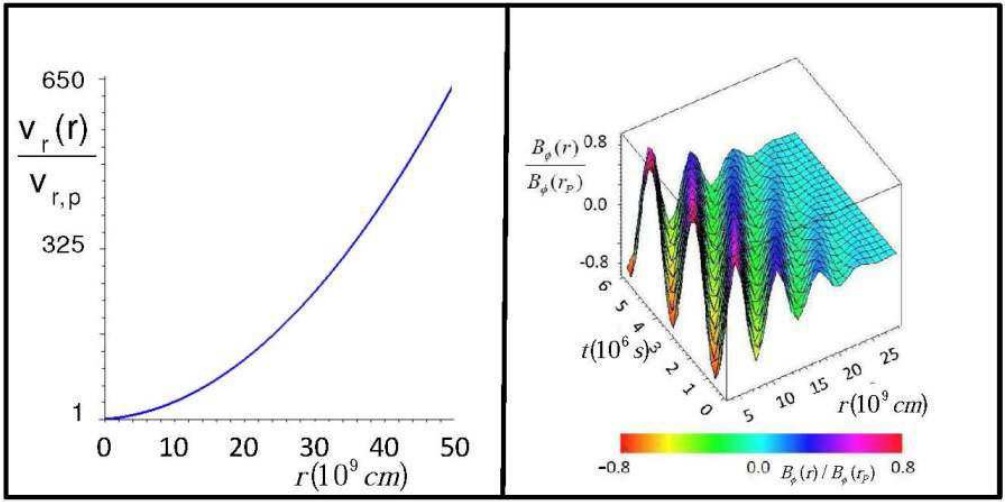}}
\caption{Left panel: the growth of the radial velocity $v_r$ (normalized to the value at the bottom) in the radiative layers of our AGB star, where $w(t)$ is chosen as a linear function of $t$. Right panel: a 3D representation of the magnetic field, as a function of time and radius, for the same layers, choosing an oscillating form for $\Phi(\xi(r,t))$, i.e. a wave-like solution for $B$. Here $\omega =2 \pi/80$ yr$^{-1}$ = 1.6 10$^{-5}$ sec$^{-1}$ was taken from \citet{n+08}. A color code for the (relative) amplitude of the magnetic fied is also given both in the right panel and in its colour bar (colors are available in the on-line version). The absolute value of the field at the level $P$ is in this case a free paramter of the model.}
\label{radiative1}
\end{figure*}

The form of the function $\xi$ is immediately suggestive of the argument of a wave-like solution. If we chose $\Phi(\xi) = A \cos(h \xi) = A \cos(hw(t)+ h/r)$, with A = const and we adopt $h = r_P$ and $w(t) = (\omega/r_P) t$, we can rewrite the relations (10) and (11) as:
$$
v_r = v_{r,P}{\Big(\frac{r}{r_p}\Big)^2} \eqno(10a)
$$
$$
B_{\varphi}(r,t) = B_{\varphi,P} \cos(\omega t + r_P/r)\Big(\frac{r_P}{r}\Big)^2\eqno(11a)
$$
where $v_{r,P}= \omega r_P$, and $B_{\varphi,P} = A r_P^{-2} $. This solution satisfies dimensional constraints on
$v_r$ and $B_{\varphi}$ if $A$ has the dimensions of a magnetic flux [units Mx] and $\omega$ is measured in sec$^{-1}$ (i.e. is a pulsation). For the sake of
comparison, we can take this last parameter from the work by \citet{n+08}, where it was shown that, for the stellar structure
discussed by BWNC, and in conditions for which a dynamo is actually sustained, $B_{\varphi}$
might have (at the base of the convective envelope) a period of the order of $T \sim 1/80 yr$ (see Figure 3 in the quoted paper). In that case
$\omega \sim$ $1.6\times 10^{-5}$ $s^{-1}$ (any other choice would obviously be possible in our general approach). Figure \ref{radiative1} shows the ensuing behavior for $v_r$ and $B_{\varphi}$. This solution describes a magnetic field that oscillates in time and space and is spatially distorted by a phase shift varying as $1/r$.

We notice that if, instead, $\Phi(\xi) = const$, then we can choose freely
$w(t)$. Even a solution for which the velocity is oscillating is viable. If we want, for example, that:
$$
v_r=\displaystyle\frac{{\rm d}w(t)}{{\rm
d}t}\, {r}^{-(k+1)}
$$
starting from:
$$
w(t) = C \cos(\omega' t + \vartheta) \eqno(11b)
$$
then we have to impose:
$$
C = - \frac{v_{r,P}}{r_P^2 \omega'}
$$
\begin{figure*}[t!]
\centerline{\includegraphics[width=0.8\textwidth]{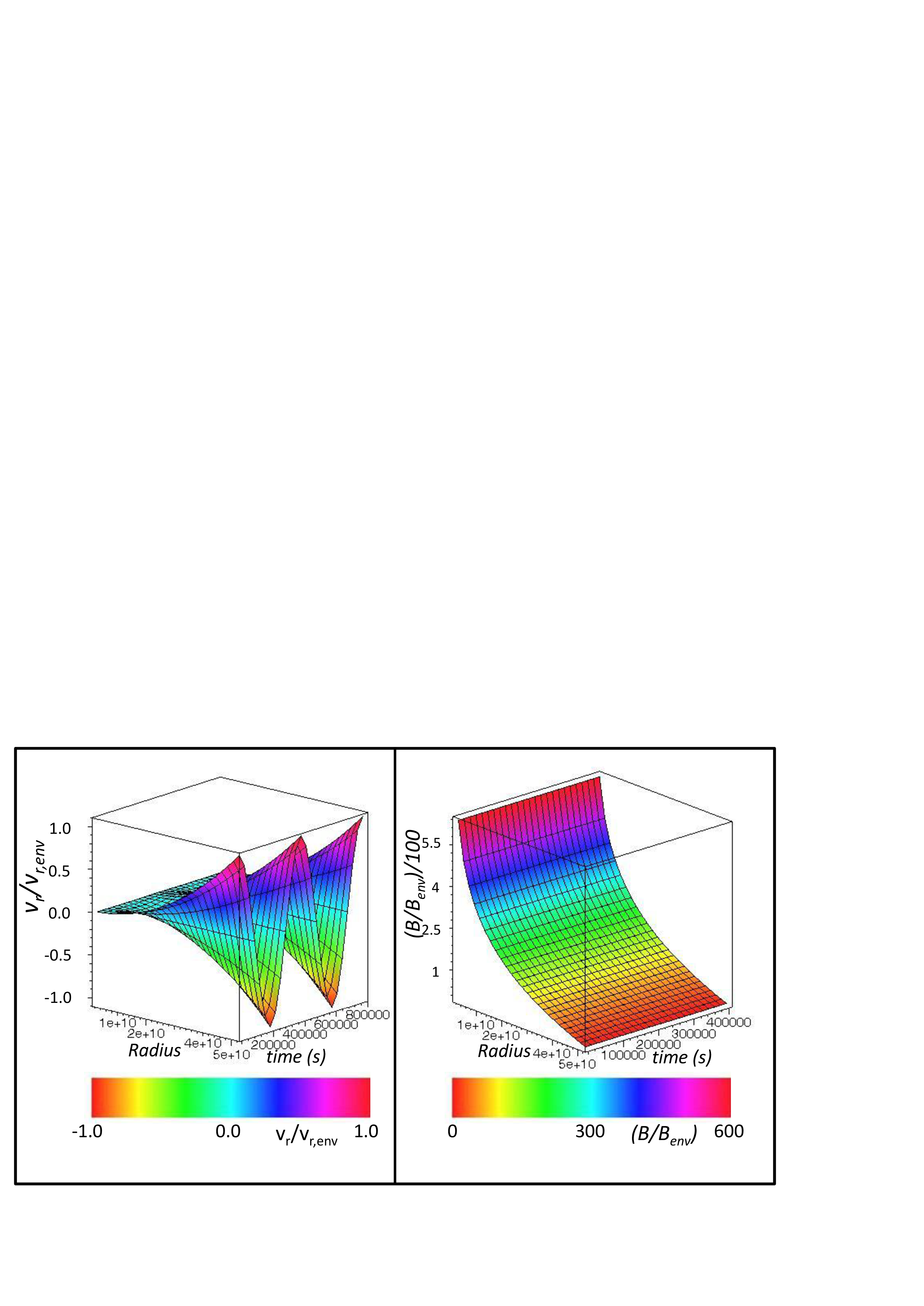}}
\caption{Left panel: a pulsational solution for $v_r$ in the radiative zones above the H-burning shell, as the one depicted here in the 3D plot and described in the text, is permitted by the MHD equations. Here $\omega =2 \pi/80$ yr$^{-1}$ = 1.6 10$^{-5}$ sec$^{-1}$ is assumed from Figure 3, for the sake of simplicity. The normalization is to the values at the envelope bottom, where the oscillation of $v_r$  is most visible. Right panel: the corresponding behavior of the magnetic field in the same layers (note that the extension of time axis is only about half that of left panel). The relative amplitudes of the radial velocity and of the field, with respect to those at the envelope base are shown both in 3D and with a color code shown in the corresponding bars (colors are available in the on-line version). As  discussed in section 6, reasonable values for the normalizing factors $v_{r,env}$ and $B_{env}$, for the case in which one considers not the average field, but confined substructures like flux tubes, are of the order of 100 m/sec and of 2$\times$10$^4$G, respectively. Hence, for flux tubes $B_P$ would be of a few 10$^6$ G, as suggested by BWNC, and the absolute value of  $v_{r,p}$ would be about 15cm/sec.}
\label{radiative2}
\end{figure*}
Then the magnetic flux is preserved. In such a case $v_{r}$ has the form displayed in Figure \ref{radiative2} (for simplicity, we adopted $\omega' = \omega$, as in Figure 3). Notice that the oscillation here is not distorted, because the phase is no longer modified by the term $\propto 1/r$. {The solution hence describes a pulsational behavior (see also secion 2.2)}. We recall that several pulsational phenomena occur in evolved stars, whose nature is not understood yet \citep{citaz}; oscillating magnetic structures might in principle be advocated also in this case.

Concerning the azimuthal component of the velocity, $v_{\varphi}$, it can be obtained from equation (A-3), which becomes:
$$
\mu r^2 \frac{\partial^2 v_{\varphi}}{\partial r^2} + r \Big(\mu - c_0 \frac{{\rm d}w(t)}{{\rm d}t}\Big) \frac{\partial v_{\varphi}}{\partial r} - \frac{c_0}{r}\frac{\partial v_{\varphi}}{\partial t} 
$$
$$
+ {v_{\varphi}}\Big(\frac{c_d c_0}{r}- c_0 \frac{{\rm d} w(t)}{{\rm d} t} -\mu\Big) = 0 \eqno(12)
$$
With $c_0= \varrho_{P} r_P^3$, all the terms of equation (12) are measured in units of (g sec$^{-2}$). Once $B_{\varphi}$ and $v_{\varphi}$ are known, the pressure can then be determined from equation (A-2), which assumes the form:
$$
\frac{{\rm d} P}{{\rm d} r} = - \frac{c_0 }{r^3}\frac{\partial \Psi}{\partial r} - \frac{1}{4\pi} \Big(\frac{\partial B_{\varphi}}{\partial r}\Big) B_{\varphi} - \frac{c_0}{r}\Big(\frac{{\rm d}^2 w(t)}{{\rm d} t^2}\Big)
$$
$$
 - 2 c_0 \Big(\frac{{\rm d} w(t)}{{\rm d} t}\Big)^2 + \Big(\frac{c_d c_0}{r}+ 3 \mu \Big)\Big(\frac{d w(t)}{dt}\Big)
$$
$$
- \frac{1}{4\pi r} B_{\varphi}^2 + \frac{c_0}{r^4}v_{\varphi}^2 \eqno (13)
$$
Obviously, for a linear choice of $w$ the term containing its second derivative would be zero. All the terms of equation (13) are measured in units of (g cm$^{-2}$ sec$^{-2}$).

We notice that in our frozen field hypothesis the radial velocity does not depend on drag and is a function of $r$ only. This descends from the form of the continuity equation in polar coordinates, from the radial dependency of $\rho$ in Figure 1 and from
the exclusion of diffusion processes that would link the velocity to thermal exchanges, hence to dissipative effects, through equation (5): see also how $v_r$ is obtained in the Appendix. Concerning $v_{\varphi}$ we warn that we cannot put too much emphasis on it. Indeed, in order to minimize the complexity of the problem for the radial velocity (where our main interest is concentrated) the
angular part of the treatment was taken very simple, neglecting any component related to stellar (differential) rotation. This simplification, for the specific layers and phases we are dealing with, is not unreasonable. Recent models indeed show that the angular velocity outside the degenerate core of an AGB star is in general small (although varying in a differential way); any rotation rate of these zones, as left behind after the Main Sequence, is in particular reduced by large factors at the AGB stage \citep[orders of magnitude: see e.g.][]{pier}. A complete treatment for $v_{\varphi}$ would require considering also latitudinal drifts and latitudinal pressure gradients \citep[see e.g.][]{remp}; hence it would actually require a more general 3D approach. Notice that such latitudinal drifts certainly exist (see Section 5); they represent a precession-like variation of the velocity vector around the radial direction, thus making helicity $\Big(H = \int {\bf v} \cdot ({\bf \nabla \times }{\bf v}) d^3r$\Big) different from zero, as it must be in dynamo mechanisms.

It is instructive to derive the form of the solution if not only diffusive processes, but also the
aerodynamic drag is negligible. In such a case it is easy to show that, mathematically, the solutions reduce to:
$$
v_r=a_1 r^{2},\quad\quad v_{\varphi}=a_2r^{-1}+a_3
r^{a_1 (r_P)^3 \varrho_{P} /\mu+1} \eqno (14)
$$
and:
$$
B_{\varphi}=r^{-2}\Phi(\xi), \quad [\xi=r^{-1}+a_1 t] \eqno (15)
$$
Here, again, the constants $a_i (i=1,3)$  and the function $\Phi(\xi)$ must be specified respecting the dimensions of $B_{\varphi}$ and $v_r$. Obviously the case $\Phi(\xi)$ = const, which implies a field that scales as the reverse of the distance squared, preserving the magnetic flux,
remains also in this case an acceptable solution (as assumed in BWNC).

\section{Buoyancy and matter deposition in the convective envelope}

For the convective envelope, excluding the innermost and outermost regions, the
bulk of the mass closely resembles a polytropic function of index 3/2. Also the
dependency of the density on radius has the form:
$$
\varrho= {c_1 r^{k}},\label{3rhoc} \eqno (16)
$$
where $k = - 3/2$, and $c_1 = {\varrho_{env}} \cdot r_{env}^{3/2}$ (see Figure \ref{prhocon}).
In this case, unfortunately, the role of analytical solutions is minimal.
Indeed, the physics is dominated by turbulence at all scales. Micro-turbulence,
as a characteristic of the plasma, can be included in MHD equations on the basis of reasonable
considereations of the microscopic behavior of particles. Macro-turbulence and the motions of the convective eddies are instead difficult to evaluate; the global
behavior of turbulence is still a challenge for physics and is certainly not approachable
analytically. This is even more so for any 3D approach. 

There is, however, a useful exercise that can be done, also in this case, with MHD. We can 
at least estimate the radial buoyancy velocity profile and the behavior of the toroidal 
field component, neglecting for the moment macro-turbulence. Then we can briefly discuss (only 
semi-quantitatively) the effects expected from convection. We remember that these last are 
twofold; convection on one side may accelerate the upward flow \citep{web}; but on the
other hand it makes magnetized structures prone to disruption phenomena
\citep{mi92,pm97}. We can discard in our preliminary analytical treatment the
temporal dependency previously analyzed in the radiative zones as, due to the
problems mentioned, our analytical solution cannot be more than a schematic
exercise. Adopting the choice of equation (16) and of Figure \ref{prhocon}, we make the same
assumptions presented elsewhere in this paper concerning the components of the velocity
vector and of the magnetic field. Then, solving again the MHD equations in 2D (as in Section 3) we derive:
$$
v_r={a_0}{r^{1/2}},\quad\quad B_{\varphi}=\frac{b_0}{r^{1/2}} \eqno (17)
$$
where $a_0 = v_{r,env} r_{env}^{-1/2}$ and $b_0 = B_{\varphi,env} r_{env}^{1/2}$. (Note that, in case time were explicitly included, the dependency of $v_r$ on $r$ would be the same). The equation for $v_{\varphi}(r)$ now becomes:
$$
\mu r^2\frac{{\rm d}^2 v_{\varphi}}{{\rm d} r^2}-r(a_0 c_1 - \mu)\frac{{\rm d} v_{\varphi}}{{\rm d} r}-
(a_0 c_1 + \mu - 
$$
$$
c_d c_1 r^{1/2})v_{\varphi}=0, \eqno (18)
$$
Equation (18) can be explicitly solved. We have two possible
solutions, depending on the presence or absence of drag:

{\bf i) With drag ${(c_d \neq 0)}$:}

$$
v_{\varphi} = r^{\displaystyle a_0 c_1/2\mu} \times F \eqno(19)
$$
with
$$
F = c_2{\rm J}\left(\frac{2(a_0 c_1 + 2\mu)}{\mu},
4\,({c_d})^{1/2}
\sqrt{\frac{c_1}{\mu}}r^{1/4}\right) + 
$$
$$
c_3{\rm Y}\left(\frac{2(a_0 c_1 + 2\mu)}{\mu},
4\,({c_d})^{1/2}\sqrt{\frac{c_1}{\mu}}r^{1/4}\right) \eqno(19a)
$$

\noindent{where $J$ and $Y$ are the Bessel's functions of the same names and $c_2, c_3$ are integration constants.} Notice that, if we indicate with $x$ the pure number $(a_0 c_1)/2\mu$ then $c_2$ and $c_3$ must be either zero or constants with units [L$^{1-x}$T$^{-1}$].
The azimuthal velocity opposes, for the Faraday's law, any original rotational
speed, so that transport induced by magnetic fields is also a promoter of
rotational braking in the convective envelope, especially in the innermost envelope layers (as the Bessel functions oscillate, but also approach zero for very large $r$).

{\bf ii) Without drag ${(c_d \simeq 0)}$:}

Here the solution simply becomes:
$$
v_{\varphi}=\frac{c'_2}{r}+c'_3r^{\displaystyle
a_0\ c_1/\mu+1} \eqno (20)
$$
Notice that we must respect the physical units for velocity. However, this is possible only if {\it either} $c'_2$ or $c'_3$ is zero. Obviously, with or without drag, the solution:
$$
\mathbf{v_{\varphi}=0}
$$
is always possible, if {\t both} $c'_2$ and $c'_3$ are zero.

Concerning the pressure, we have:
$$
\frac{{\rm d} P}{{\rm d} r} = -\frac{c_1}{r^{3/2}}\frac{{\rm d} \Psi}{{\rm d} r} + \frac{1}{8 \pi r^3} \times G \eqno(21)
$$
with:
$$
G = - 2 \pi a_0 r^{3/2} (2a_0 + 3\mu) + 
$$
$$
+ 8 \pi c_1 (r^{1/2} v_{\varphi}^2 + a_0 c_d r^2) -b_0^2 r 
\eqno(21a)
$$
From (21) on gets in any case that $P$ will not go formally to zero at large distances. The physical constraints say that the gas pressure and the gravitational potential do. The rest is accounted for only by magnetic effects. Hence, a non-vanishing pressure at the surface would simply mean that, if expanding fields were to survive up to outer stellar layers (but see next Section), then a magnetic wind would result.

\begin{figure}
\centerline{\includegraphics[width=\columnwidth]{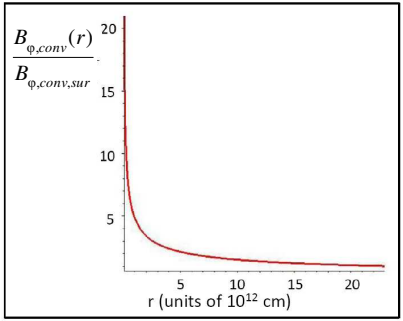}}
\caption{The dependency of the toroidal magnetic field $B_{\varphi}$ on radius in the convective layers, discarding any effect of macro-turbulence.}
\label{vBconv}
\end{figure}

\begin{figure*}[t!!]
\centerline{\includegraphics[width=0.7\textwidth]{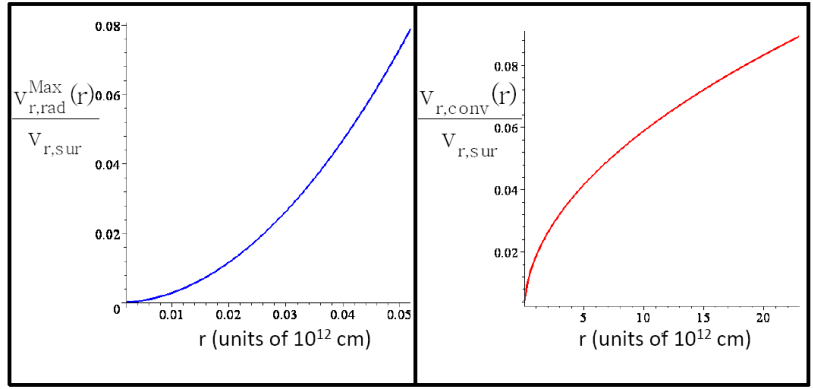}}
\caption{The dependency of the buoyancy velocity upon radius in the radiative layers
(left panel) and in the convective layers (right panel). The two plots show the transition
from a small region with a fast (parabolic) growth, to a much larger convective envelope with a slower (square root) growth. In this last zone the effects of macro-turbulence are not included.}
\label{vrglobal}
\end{figure*}

Figure \ref{vBconv} shows the profile $B_{\varphi}/B_{\varphi,sur}$ in the convective region (the behavior is obviously the same for the average field and for any possible organization of it in structures, like flux tubes). As here we have simplified to the minimum our treatment, suppressing
time dependency, the plot has to be seen only as an illustrating schematic behavior, in the absence of macro turbulence.

\section{A 3D approach for the velocity in radiative and convective layers}

Very often, 2D solutions in a problem with a complex geometry is at risk of not being general enough to give a realistic representation of the real physical situation. We can make  for this a simple example. Let's consider our radiative zones, and a thin circular flux ring in them, under a 3D geometry. Let the ring be in the equatorial plane and with the center in the center of the star. If $\rho= \gamma r^k$ and if the azimuthal and meridional components of the velocity are both negligible, then the continuity equation seems to imply that $v_r \propto r^{(-k-2)}$. In the radiative 
layers we showed that $k = -3$, hence this requires $v_r \propto r$, at odds with the solution (10a) of the equations provided in Section 3.
 
This contradiction may appear, in this specific case, misleading, as: i) $v_{\varphi}$ is not negligible (see equation 12); and ii) in the example above we are implicitly assuming spherical geometry (pure dependency on $r$), which can never be the case with a magnetized plasma, as the magnetic field is solenoidal and we would violate the Gauss' theorem. Nevertheless, the example is disturbing enough to induce doubts 
on the generality of our solution and to require a demonstration, made in a 3D framework, that our results implying $v_r \propto r^2$ remain valid also in that case.

In general, this demonstration would be very difficult analytically, but the physical situation prevailing in the radiative layers of an AGB star below the envelope is very special, as we mentioned, and is subject 
to a quasi-ideal MHD treatment. In this case the radial and meridional velocity do not depend on the whole, very complex, system of equations from (1) to (5), but only on the continuity equation (1), which (in 3D, i.e. introducing also the mentioned meridional velocity $v_{\vartheta}$) becomes:
$$
\frac{\partial \varrho}{\partial t}+\frac{1}{r^2}\frac{\partial (\varrho v_r r^2)}{\partial r}
+\frac{1}{r \sin \vartheta}\frac{\partial (\varrho v_{\vartheta} \sin \vartheta)}{\partial \vartheta}+
$$
$$
\frac{1}{r \sin \vartheta}\frac{\partial (\varrho v_{\varphi})}{\partial \varphi} = 0 \eqno(22)
$$
Also in these conditions, we can continue to impose that any density shift due to magnetic fields is small, so that equation (9) continues to hold. Then one obtains easily:
$$
r \frac{\partial v_r}{\partial r} + \frac{\partial v_{\vartheta}}{\partial \vartheta} + \frac{1}{\sin \vartheta}\frac{\partial v_\varphi}{\partial \varphi}+ \cot \vartheta v_{\vartheta} - v_r = 0 \eqno(22a)
$$
For simplicity, we further assume that the equatorial section of the star remains circular, so that $v_{\varphi}$ 
does not depend on $\varphi$ and the corresponding derivative is zero. Hence the full 3D approach simplifies to a
double-2D one (the first, in the ($r$, $\varphi$) plane, already performed; the second, in the ($r$, $\vartheta$) plane, presented in this section).  For the rest, we must identify functions $v_{r}$ and $v_{\vartheta}$ (which will be in general functions of both $r$ and $\vartheta$), satisfying 
equation (22a). Notice that we have a single relation linking two variables, so that out of the infinite number of possible solutions  one should look for those satisfying our physical problem.

In order for the 3D approach to confirm our previous 2D solution, the equatorial value of the radial velocity $v_{r,eq}$ must differ from the 3D one (averaged over $\vartheta$) only by some numerical factor. From the physical point of view we can speculate  that the radial velocity itself should show lager values toward the equator or at some intermediate latitudes than toward the poles, in order to describe a toroidal, or double-toroidal configuration, as seen in active stars. (This is a rather weak way of posing boundary conditions, but on the actual geometry and on the internal dynamical structure of an AGB star we know virtually nothing). We must also require that $v_{\vartheta}$ be zero at the equator, in order to reproduce the 2D solution of equation (10a). 

With the above constraints, we can discuss equation (22a) starting from the simplest possible assumptions. For 
example, it easy to see that, if $v_{r}$ does not depend on $\vartheta$, a solution $v_r \propto r^2$ does not 
hold, and actually  $v_r \propto r$. This says that the example made at the beginning of this Section remains, in
principle, even when $v_\vartheta$ and $v_\varphi$ are not negligible, provided that
$v_r$ does not depend on them. As said, this is however incompatible with the solenoidal nature of magnetic fields. 
In order to have a chance that a relation $v_r \propto r^2$ be valid, the same 
$v_r$ must also vary with $\vartheta$. In accordance with the conditions posed above on $v_{\vartheta}$, one can, as an example, choose: $v_{\vartheta} = - f(r) \sin 2{\vartheta}$, looking for an expression $v_r = g(\vartheta) r^2$, with the condition that at the equator ($\vartheta =0$) $g(0) = v_{r,P}/(r_P)^2$.
Substituting this tentative form into equation (22a) one finds a simple solution  satisfying the continuity equation and also our basic physical requests. 
In particular we have:
$$
v_r(r,\vartheta) = {\frac{1}{2}}{\frac{v_{r,P}}{{r_P}^2}}r^2 \left[ 2 \cos^2 \vartheta - \sin^2 \vartheta \right] \eqno(23a)
$$
$$
v_{\vartheta}(r,\vartheta) = - {\frac{1}{4}}{\frac{v_{r,P}}{{r_P}^2}} r^2 \sin 2\vartheta \eqno(24a)
$$
Notice that the third component $v_{\varphi}$ enters the continuity equation only through the derivative of ($\rho v_{\varphi}$) with respect to ${\varphi}$, which is zero in our hypotheses. It will depend on magnetic fields in a complex way: even in 2D, equation (12) said that. However, the simple form of MHD holding in our case makes the behavior of $v_{\varphi}$ irrelevant for $v_r$ and $v_{\vartheta}$, so that we can discuss our generalization that includes meridional motions without the need of invoking the explicit form of the azimuthal velocity. What is important for us is to verify if it is true that $v_r$ is proportional to $r^2$. We can deduce this, in two or in three dimensions, without referring directly to $v_{\varphi}$. This also tells that any form of differential rotation possibly descending by an exact derivation of this last function would not alter our basic conclusions for the expansion.

At the equator, where $v_{\vartheta}$ vanishes, the solution for $v_r$ reduces to the one derived in the 2D case. Also its average over the latitude (from $-\pi/2$ to $\pi/2$) differs from our solution (10a) only by a a numerical factor $\alpha$ (with $\alpha$ = 0.25). This is obviously irrelevant: it can be included in the multiplying coefficient of equation (10a), which depends on the free parameter ${v_{r,p}}$. Notice that $v_{\vartheta}$ vanishes also at the poles.

\begin{figure*}[t!!]
\centerline{\includegraphics[width=0.35\textwidth]{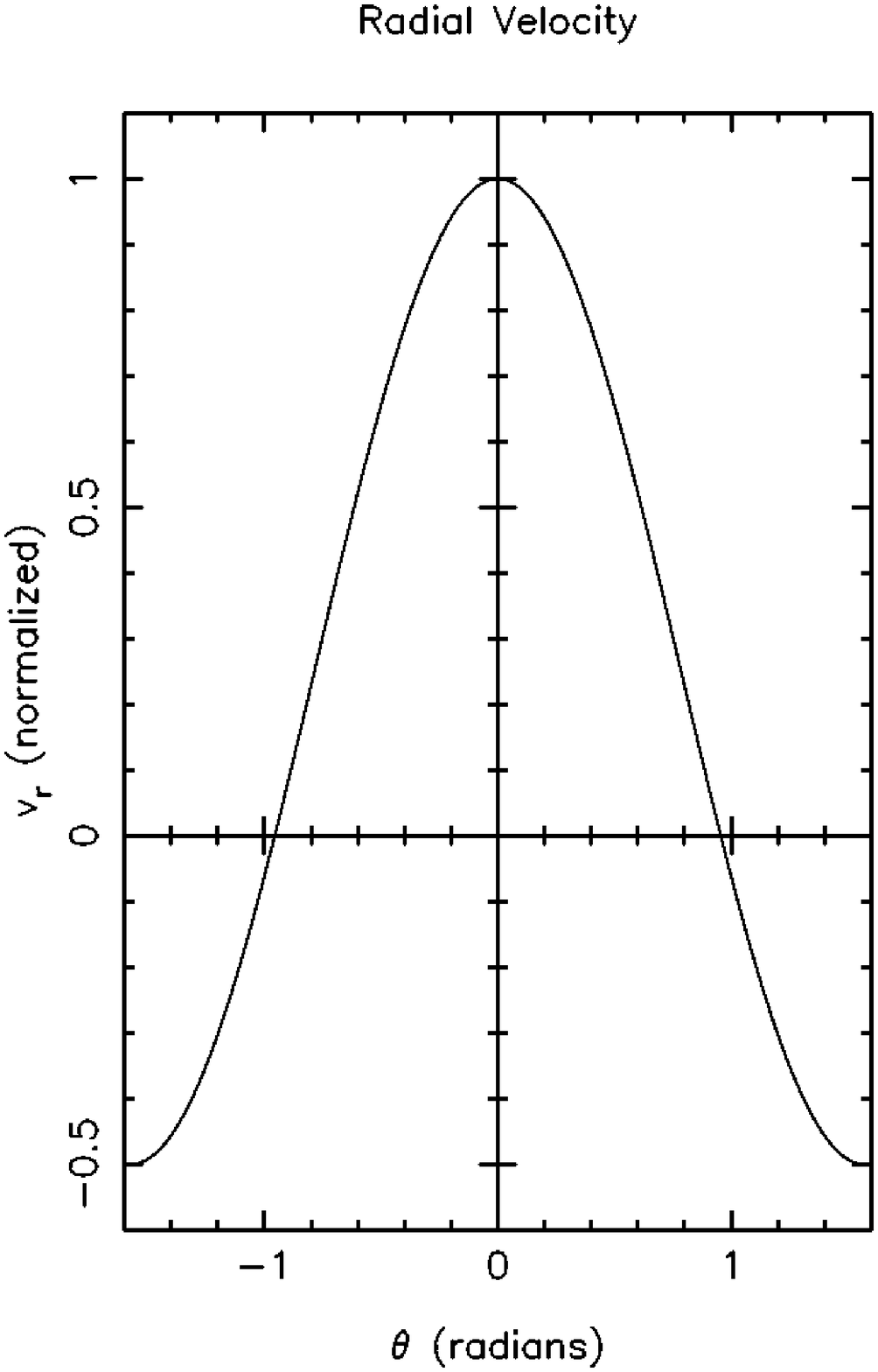}
\includegraphics[width=0.35\textwidth]{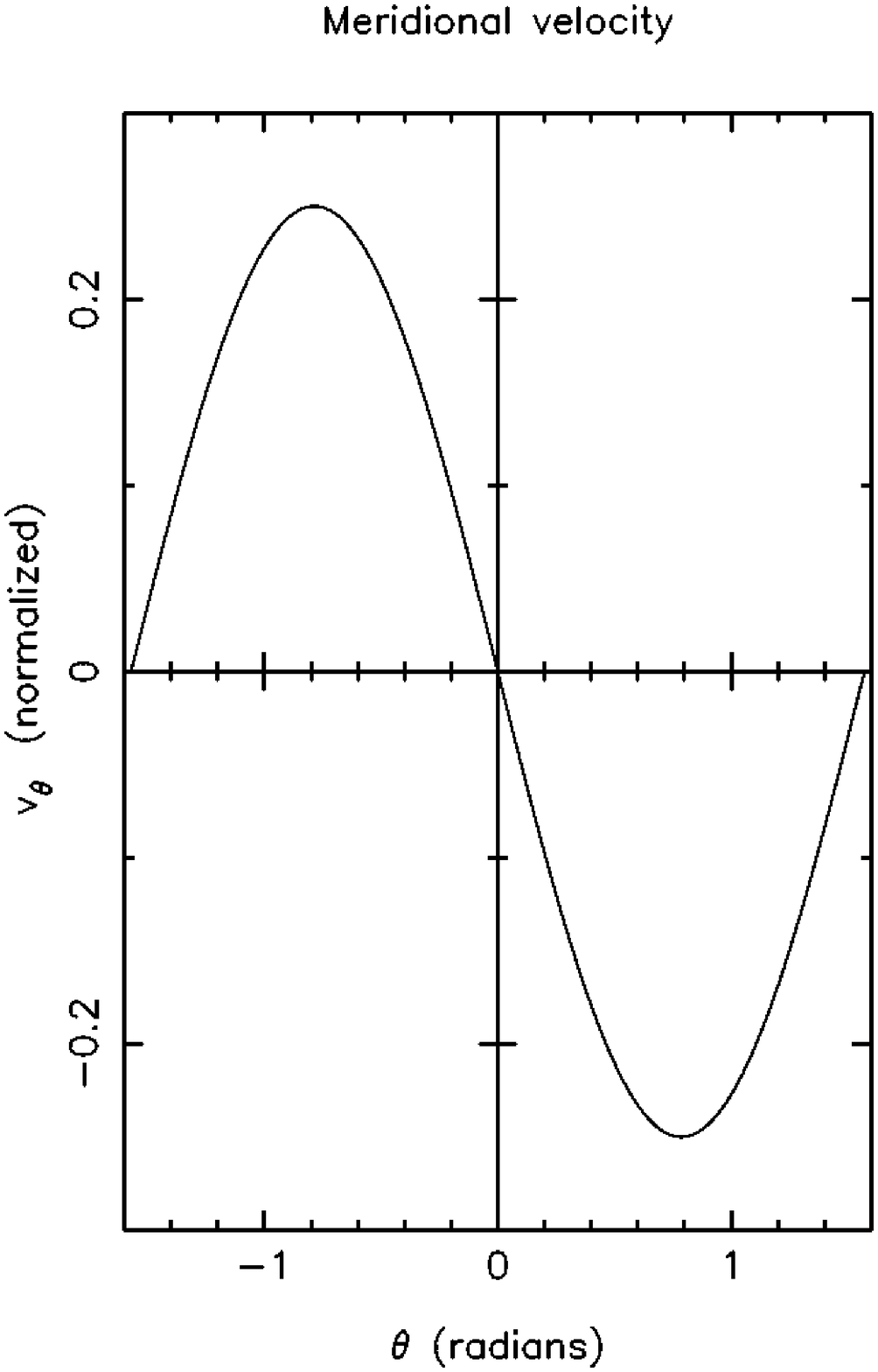}}
\caption{The dependency  of the radial (panel a) and latitudinal (panel b) 
components of the velocity as a function of $\vartheta$ and of $r$ in the first 3D solution discussed in the text: see equations (23a) and (24a). In the plots the velocity is normalized, dividing by the  factor $v_0 = {v_{r,p}}/{r_p^2} \times r^2$, in order to show only the dependency on latitude.}
\label{vel2d1}
\end{figure*}

\begin{figure*}[t!!]
\centerline{\includegraphics[width=0.35\textwidth]{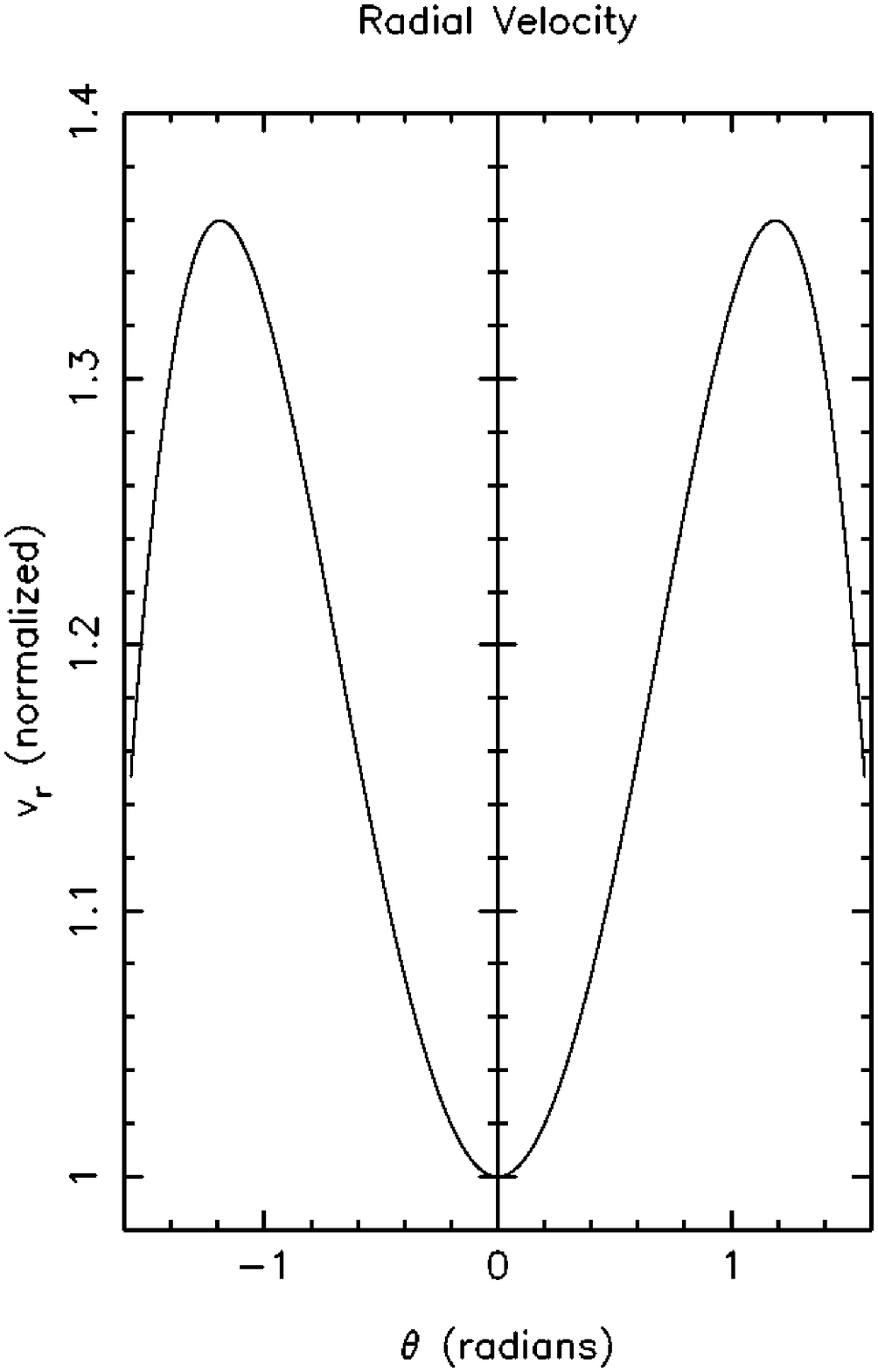}
\includegraphics[width=0.35\textwidth]{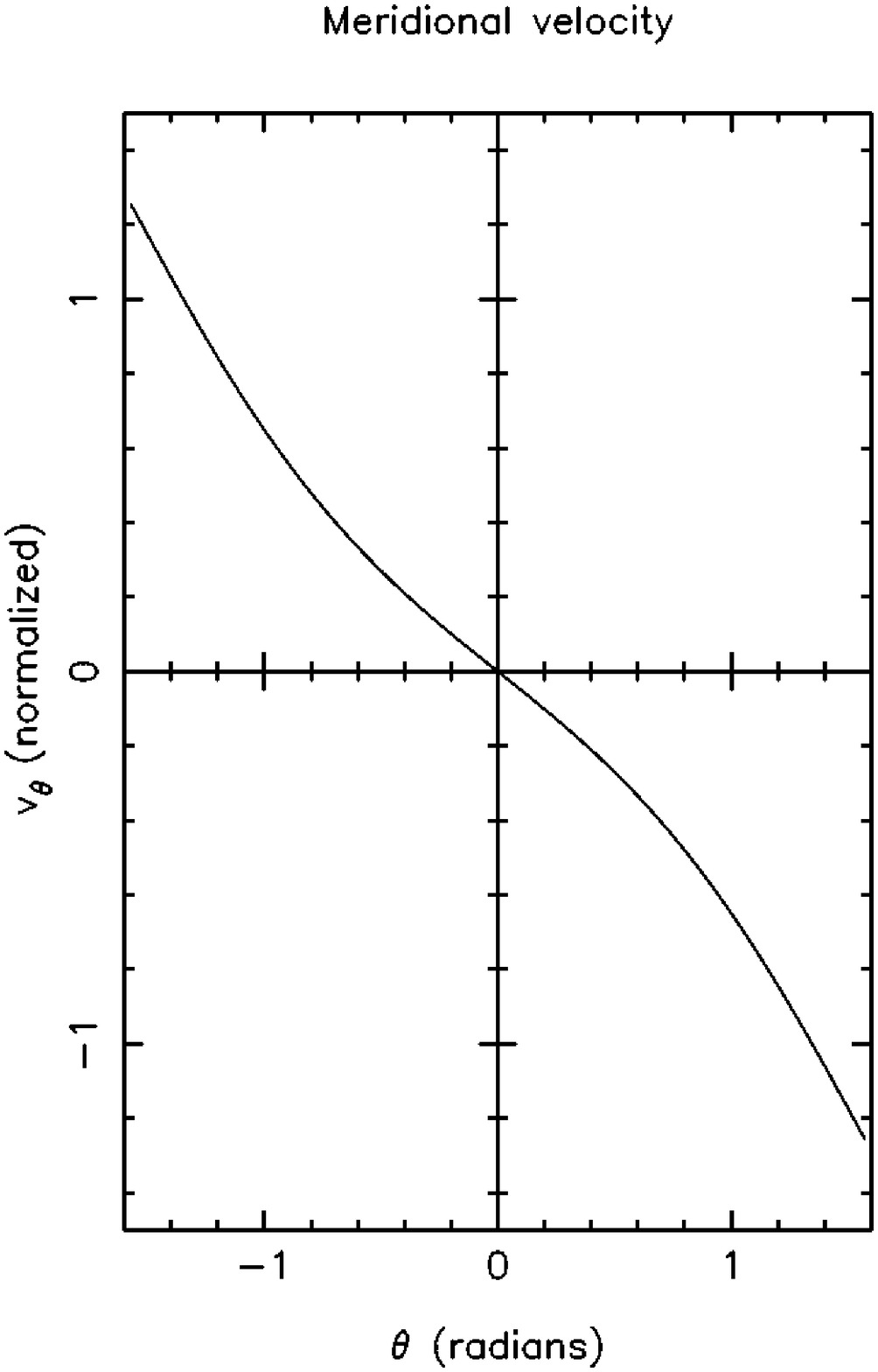}}
\caption{The dependency  of the radial (panel a) and latitudinal (panel b) 8
components of  the velocity as a function of $\vartheta$ and of $r$ in the second 3D solution discussed in the text: see equations (23b) and (24b). In the plots the velocity is normalized, dividing by the  factor $v_0 = {v_{r,p}}/{r_p^2} \times r^2$, in order to show only the dependency on latitude.}
\label{vel2d2}
\end{figure*}

\begin{figure*}[t!!]
\centerline{\includegraphics[width=\textwidth]{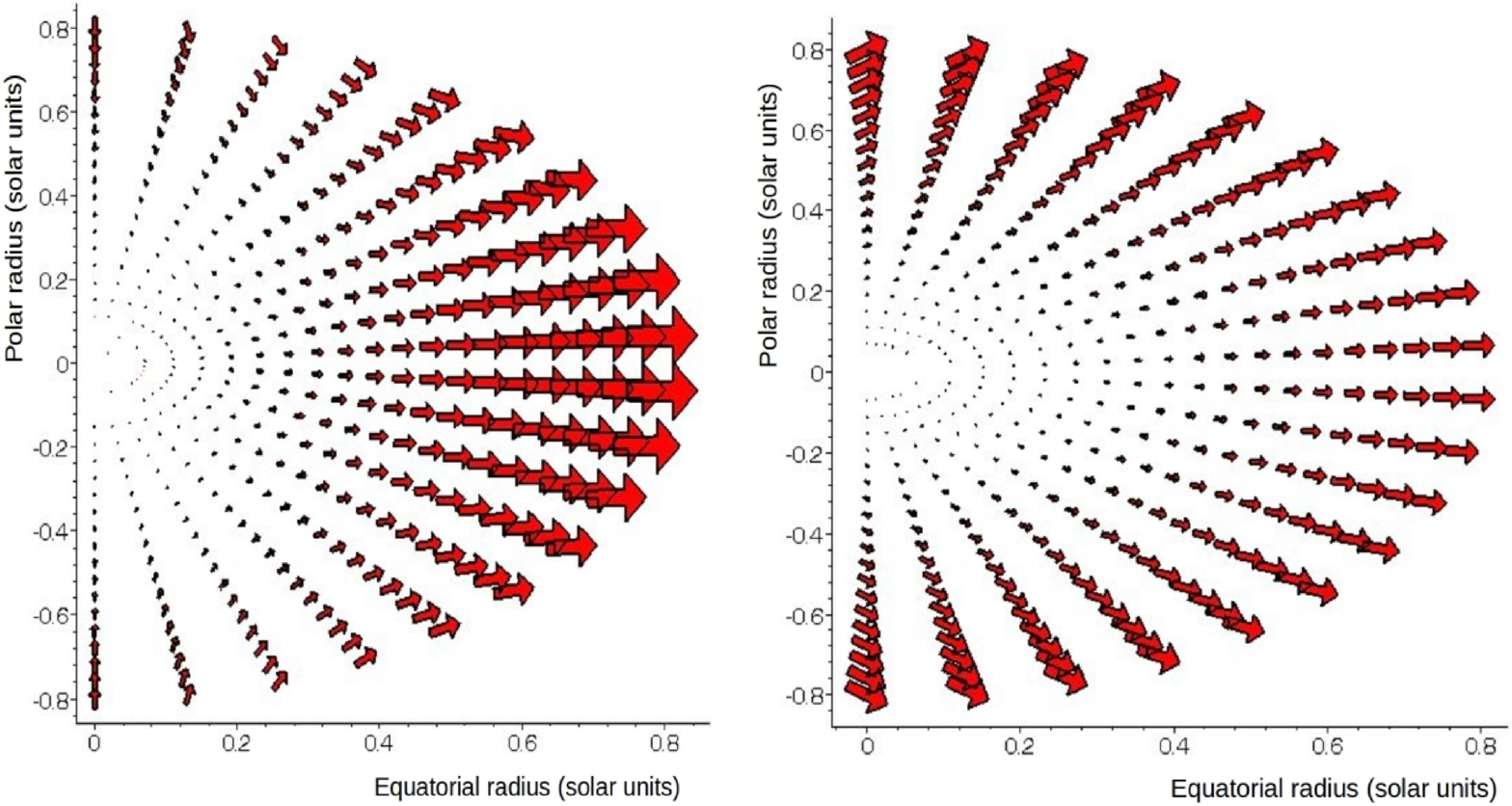}}
\caption{Illustrations of the velocity field over a stellar hemisphere: the left panel combines solutions (23) and (24), the right panel solutions (23a) and (24a). See text for explanations. The axes show the equatorial and polar extension of the radiative layer, in units of the solar radius.}
\label{velarr}
\end{figure*}

A graphic representation of the forms (23a) and (24a) is presented in Figure \ref{vel2d1}. The behavior of $v_r$ describes a single toroidal structure, with an expansion that has its maximum radial velocity at the equator and changes sign toward the poles, implying there a very weak inflow, thus producing an annular structure. Inspection of the resulting vectorial velocity in cartesian coordinates (Figure \ref{velarr}, left panel) makes clear that the outflow largely prevails. In the figure we see a hemisphere of the star,
with the north and south poles at the extremes of the vertical axis and with the horizontal axis on the equator. The axes extend over the thickness of the radiative region (slightly less than 1 solar radius). The size of the arrows is proportional to the ratio of the velocity field strength at the given point to the average field strength at all points in the grid. In general, for its properties of symmetry and 
for the relation of its mean value with the one provided by equation (10a), this solution is a 3D extension of our previous 2D case and confirms its proportionality to $r^2$.

Another, a bit more sophisticated, solution (describing in this case a double toroidal structure, with a radial velocity that achieves its maxima at intermediate latitudes, north and south of the equator, like for the Sun and many active stars) is for 
example provided by the following relation for $v_r$: 
$$
v_r = \frac{1}{2}\frac{v_{r,p}}{r_p^2}r^2(2 \cos\vartheta \cosh \vartheta+\sin \vartheta \sinh \vartheta) \eqno(23b)
$$
In order to satisfy the continuity equation this requires that:
$$
v_{\vartheta} = - \frac{1}{2}\frac{v_{r,p}}{r_p^2}r^2 \cosh \vartheta \sin \vartheta \eqno(24b)
$$
Again, the above formulae fulfil the requirements at $\vartheta$ = 0, where they 
reduce to our 2D solution found before, because there 
$v_{\vartheta}$ = 0 and $v_r = ({v_{r,p}}/{r_p^2}) \times r^2$, which is again formula (10a). The velocity $v_r$ is again the product of two terms, depending
one on $r$ and the second on $\vartheta$; this second function (see equation 23b) does not diverge in the interval of interest (from $-\pi/2$ to +$\pi/2$, i.e. 
from the south to the north pole). It is symmetric with respect to the equator and its average corresponds to the form (10a) multiplied by a factor $\alpha$ (which is, in this case, $\alpha$ = 1.198). 

\begin{figure*}[t!!]
\centerline{\includegraphics[width=0.8\textwidth]{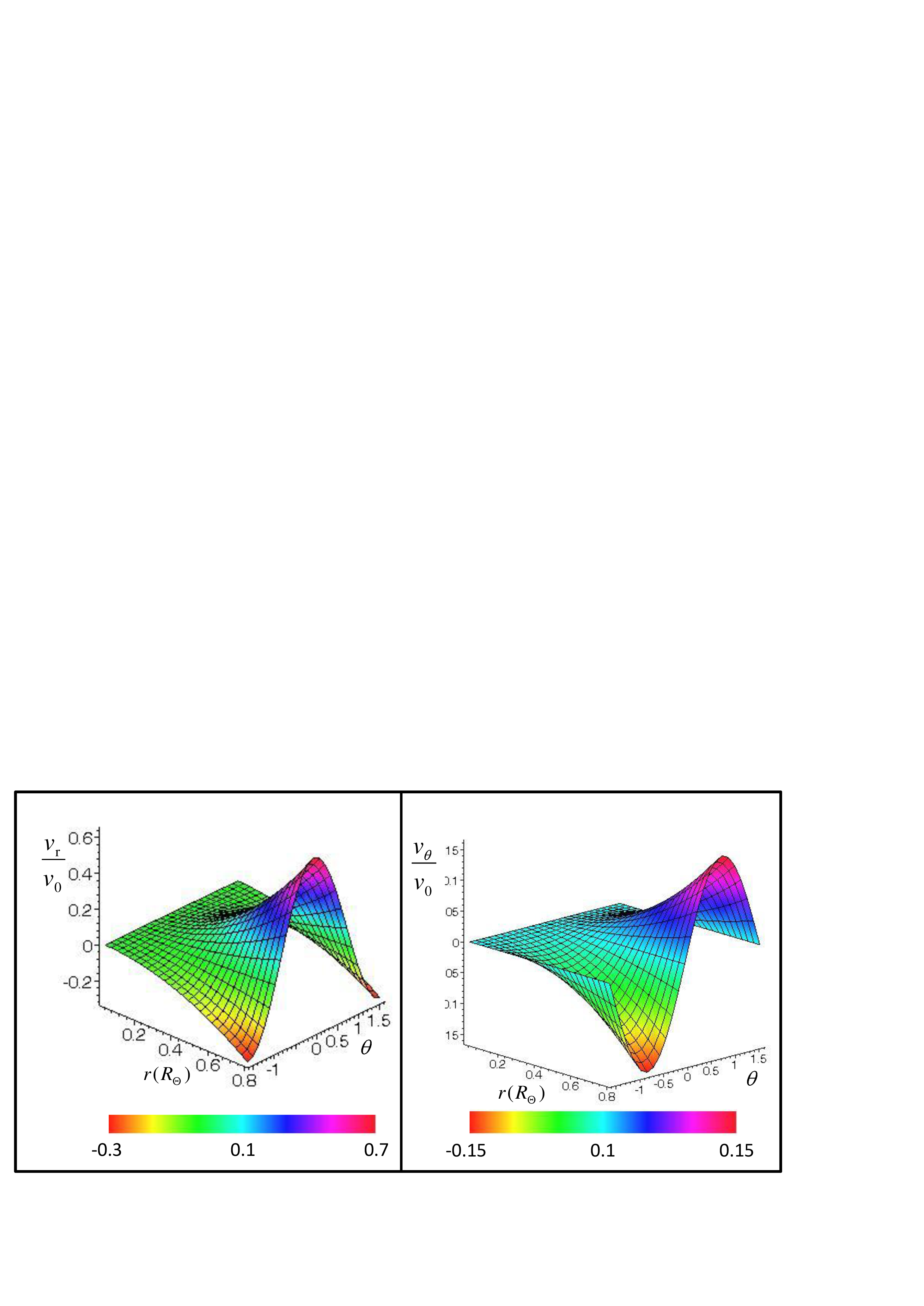}}
\caption{The dependency of the radial (panel a) and latitudinal (panel b) 
components of the velocity as a function of $\vartheta$ and of $r$ in a 3D representation of the solution of equations (23a) and (24a). In the plots the velocity is normalized, dividing by the  factor $v_0 = {v_{r,p}}/{r_p^2} \times r^2$, in order to show only the dependency on latitude. The relative amplitudes thus obtained are color-coded in the bars below the graphs. Colors are available in the on-line version.)}
\label{vel3d1}
\end{figure*}

\begin{figure*}
\centerline{\includegraphics[width=0.8\textwidth]{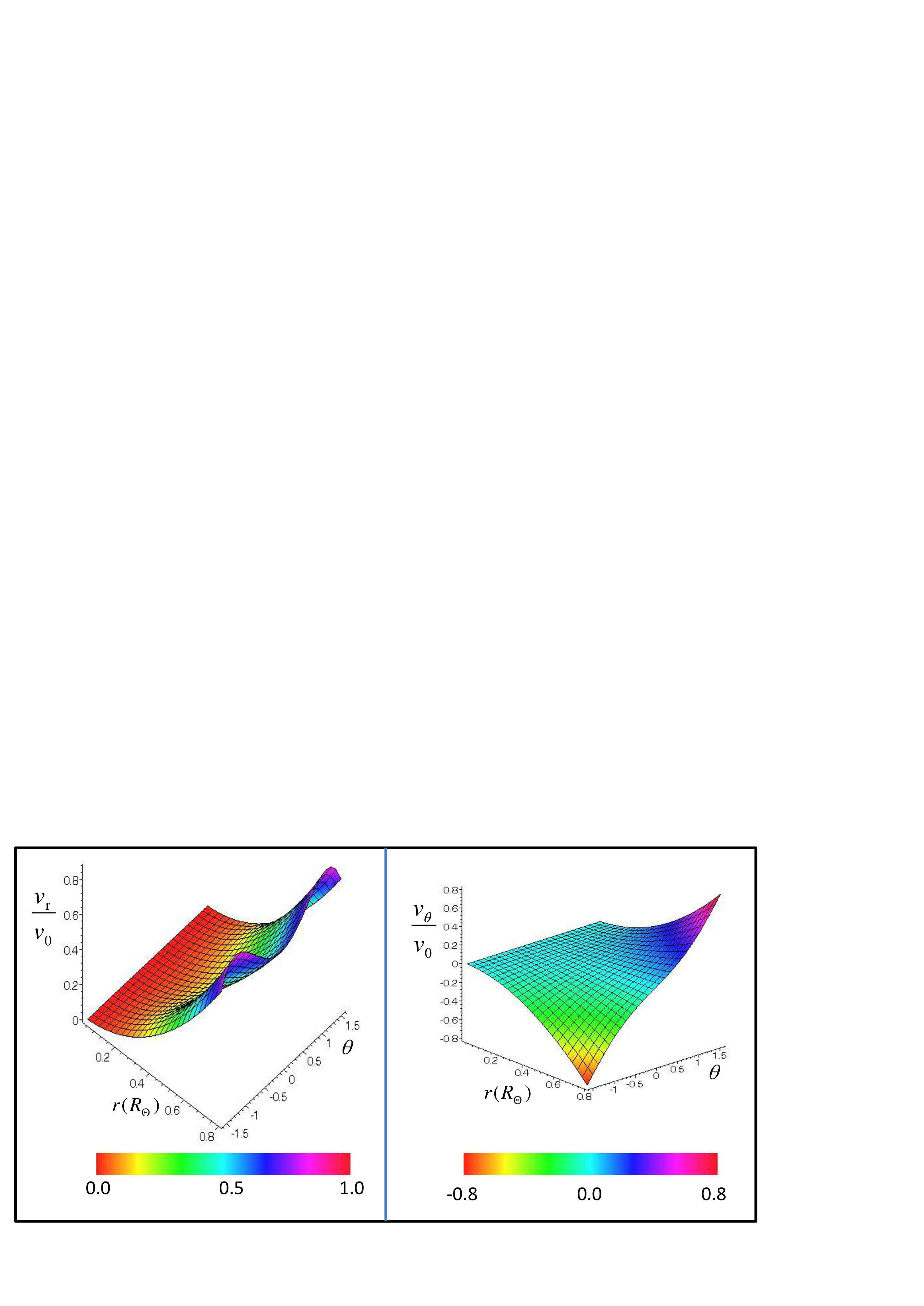}}
\caption{The dependency of the radial (panel a) and latitudinal (panel b) 
components of the velocity as a function of $\vartheta$ and of $r$ in a 3D representation of the solution of equations (23b) and (24b). In the plots the velocity is normalized, dividing by the  factor $v_0 = {v_{r,p}}/{r_p^2} \times r^2$, in order to show only the dependency on latitude. The relative amplitudes thus obtained are color-coded in the bars below the graphs (colors are available in the on-line version).}
\label{vel3d2}
\end{figure*}

Graphic representations of the forms (23b) and (24b) for $v_r$ and $v_{\vartheta}$ are presented in Figure \ref{vel2d2}. The vectorial field of the velocity is shown in Figure \ref{velarr} (right panel).
Three dimensional plots for both sets of solutions are also given in Figures \ref{vel3d1} and \ref{vel3d2}.

The above figures show that the geometry of the expansion process is not spherical: there are preferential zones in which buoyancy occurs faster, centred at the equator or at intermediate latitudes. In AGB stars, the overlying convective envelope is huge and its turbulent mixing will take care of spreading everywhere the transported material, so that what really counts is not the single or double toroidal structure that one can infer, but the fact that the material crosses the radiative zone with a velocity that 
is {\it on average} proportional to $r^2$, as in the 2D solution obtained previously.

The same verification of the 2D solution we have just presented for the radiative zones is necessary also for the convective envelope, but in this case all the cautions already advanced in 2D must be underlined again. Our approach in quasi-free MHD condition will not be really meaningful for the turbulent envelope, if not as a limiting solution: the motion must be certainly slower than we can find (due to dissipation by turbulence) and any analytical solution provides only upper limits for the expansion velocity.

Once the above warnings have been advanced, it is in any case interesting to notice that the only change we introduce in our assumptions across the bottom envelope border concerns the density (which passes from a decrease proportional to $r^{-3}$
to one proportional to $r^{-3/2}$, but still does not depend on 
${\vartheta}$). As we have given exemplifying solutions for $v_r$ and $v_{\vartheta}$ where the dependency on $r$ and ${\vartheta}$ are in separate functions, then {\it the same solutions found in this Section for the angular part of the velocity components in the radiative zones
 holds also in the convective envelope}. Hence the validation done so far for the radiative layers applies equally well to the envelope and again the general case in which latitude is considered provides solutions whose equatorial sections are the 2D solutions presented in Section 4. The only change with respect to what has been done for the radiative zone will concern the coefficient, in order to maintain continuity of the velocity across the envelope border, so that, for the envelope, we have, e.g.:
$$
v_r(r,\vartheta) = \frac{1}{2}\frac{v_{r,env}}{r_{env}^{1/2}}r^{1/2} \left[ 2 \cos^2 \vartheta - \sin^2 \vartheta \right] \eqno(23c)
$$ 
$$
v_{\vartheta}(r,\vartheta) = - \frac{1}{4}\frac{v_{r,env}}{r_{env}^{1/2}}r^{1/2} \sin 2\vartheta \eqno(24c)
$$
Similar considerations are valid for the form (23b) and (24b). 

Concerning the meridional velocity $v_{\vartheta}$, comparisons between the forms (24a) and (24b), together with Figure \ref{velarr}, illustrate well the kind of $physical$ difficulties one encounters in pushing forward a 3D modelling, independently of any mathematical intricacy. The two forms (24a) and (24b) are indeed very different, especially near the poles: the first one vanishes, while the second one does not (it is actually large). The first behavior probably does not create problems, but is also the second one realistic? It describes a latitudinal motion over-imposed to buoyancy, which is quite different from the classical meridional circulation of the the Eddington-Sweet effect, induced by rotational  distortions of the isothermal surfaces \citep{vz24}. 

The answer to the above question is forcedly: ``we do not know". On one side, the physical complexity of the huge convective envelope prevents us from developing a real model for the MHD structures surviving in it. On the other hand, no observations exist on the meridional motion of magnetic structures at the surface of AGB stars. We notice that any field component would induce new currents, hence new velocity components and new fields, in a very intricated pattern, for which we do not have any constraint. We can only say that the behavior described by (23b) and (24b) cannot be shown to be impossible on the basis of present knowledge and we are forced to stop here. In such conditions, any complete 3D model, involving also the MHD equations and not only the continuity equation, is bound to produce (if analytically feasible) extremely complex solutions for the field $B$, whose physical plausibility cannot be proven or dis-proven on observational grounds. We must therefore limit ourselves, by saying that a dedicated 3D verification shows how the 2D solution we found for the expansion velocity is reasonable and can remain valid in 3D, without the possibility of deriving much further insight.

This paragraph should in any case demonstrate that, if we take into account the non-spherical geometry necessarily established in presence of solenoidal magnetic fields, then our 2D solution for $v_r$ can be considered as a planar section of a 3D (toroidal-like) configuration. The simpler 2D form for the radial velocity can be compared to a variety of 3D solutions (that display maxima at some low or intermediate latitudes). The 2D form turns out to differ from these 3D solutions, averaged over $\vartheta$, only by a numerical factor. The non-spherical geometry is of no real consequence for mixing. Although the same solutions found for the radiative layers would remain valid for the convective envelope when assuming quasi-ideal MHD conditions, in the reality this simplification is unreasonable in presence of turbulent convection. In the convective layers any effect produced by the geometrical pattern established in the radiative zones would probably disappear, guaranteeing that mixing occurs over the whole envelope; hence our 2D solution for the radiative zone can be actually retained, without any real loss of generality.

We can therefore conclude our discussion on this point by confirming that, in physical situations like those encountered in the radiative zones of AGB stars, magnetic buoyancy can provide a radial transport velocity growing (once averaged over the latitude) as the second power of the radius. In the overlying convective envelope the growth will proceed at most with a dependency on the square root of the radius, but most probably every magnetic structure will be destroyed there (see Section 6). The general behavior of buoyant structures found in a simple 2D treatment, is now verified to hold also in 3D, when our (very limited) set of boundary conditions is taken into account. We believe this puts our suggestions on sufficiently safe grounds to accept the existence of MHD-mixing as an important possibility to consider. 

\section{Discussion and Conclusions}
In general, the relations shown in the previous paragraphs tell us that, once emerged in the convective envelope due to a fast buoyancy from the radiative layers, the material advected by the magnetic structures would have a velocity increasing more slowly with radius, even in the (non realistic) hypothesis that any disrupting effect from macro-turbulence can be ignored (see Figure \ref{vrglobal}).

We may essentially see the velocity components $v_{\varphi}$ (and $v_{\vartheta}$) as tools for
trapping part of the flux in azimuthal (and meridional) motions, from which the material would
eventually diffuse into the envelope. The above behavior depicts
an a priori promising picture for buoyant magnetic structures as drivers of
deep mixing in stars.

In order to better understand this point, let's fix the boundary conditions to
some roughly realistic values. First of all, surface average magnetic fields on
AGB stars can be derived from recent measurements in circumstellar maser
sources \citep{her,vlem}. An average value around 3G was inferred by
\citet{her}, within a spread covering the range from 0 to 20G. Let us choose
the average value (3G), just for the sake of exemplifying. A rough upper limit to the radial velocity from pure magnetic effects is fixed by the Alfv\'en velocity  $v_A$. With the choice just made for
the field, and for the stellar parameters of Table 1, this gives a final radial velocity at the surface $v_{r,sur,av} < 1.69 \cdot 10^4$ cm/sec. Should the field at the surface, $B_{sur}$ present strongly confined components (in highly magnetized flux tubes or $\Omega$-shaped loops, as in the Sun), then equation (8) should be used to determine the magnetic field in the tubes, $B_t$. In order to do this, we need an estimate for the plasma $\beta$ parameter. This can be written as:
$$
\beta =  8 \pi k_B n T/B_{av}^2 = 0.035 n_9 T_6/(B_{10})^2 \eqno(25)
$$
\citep{gary}, where $n_9$ is the number density in units of 10$^9$ cm$^{-3}$,
$T_6$ is the temperature in MK, and $B_{10}$ is the field expressed as
multiples of 10G (i.e. in our case it is 0.3 for the average surface field). At the
surface of the AGB star discussed by BWNC, $n_9$ = 1.5$\cdot 10^6$, $T_6$ =
0.003, hence we get $\beta = $ 1750. This gives an average value for the field
in substructures (tubes, loops) of $B_{sur,t} \simeq $  1255G at the surface (the
corresponding condition that the radial velocity of the tubes be lower than the
Alfv\'en velocity for them would then be $v_{r,sur,t} \lesssim 70$ Km/sec).  As,
in our solutions for the convective envelope, the velocity and the field are
proportional and inversely proportional, respectively, to the square root of the
radius, and as $\sqrt{r_{sur}/r_{env}}
\simeq $ 21.14, we have immediately, at the base of the envelope: $v_{r,env,av} \lesssim $799 cm/sec,
$v_{r,env,t} \lesssim $ 3.3 Km/sec; $B_{\varphi,env,av}$ =
63.4 G, $B_{\varphi,env,t}$ = 2.65$\cdot$10$^4$ G. This last value is larger than
the (order-of-magnitude) estimate by BWNC by less than a factor of 3 and
essentially identical to the typical field inferred for flux tubes at the base
of the solar convective envelope \citep{fan06,sol}. Should we adopt the
Alfv\'en velocity as a proxy for the buoyancy velocity, then we could
immediately derive the crossing time of the convective region, $\Delta t_c$:
$$
\Delta t_c = \sqrt{\frac{r_{env}}{v_{r,env}}} \times \int^{r_{sur}}_{r_{env}} \frac{dr}{\sqrt{r}} \eqno(26)
$$
This provides $\Delta t_c \simeq$ 0.2 yr for the flux tubes, or $\Delta t_c
\simeq$ 87 yr for the average field. However these values are only lower limits
(obtained for the upper velocity limit $v_A$).

We cannot derive a more realistic estimate for the buoyancy velocity of flux
tubes $v_{r,t}$ in convective conditions if not by taking it from the
discussions made by other authors, with approximate treatments. For example, in
\citet{vish} it was obtained that in the most internal layers of the envelope $v_{r,t}$ is
roughly $\simeq$ 0.2$v_c$, where $v_c$ is the average local convective velocity. In
the model referred to in Table 1,  $v_c$ rapidly grows, at the inner envelope
border, from zero to about 0.5 km/sec, subsequently increasing more slowly.
Hence we can assume $v_{r,env} \simeq$  0.2$\times$ 500 = 100m/sec for flux tubes. This choice is
actually much smaller than the upper limit estimated above at the base of the
envelope (3.3 Km/sec). Adopting this new choice as a boundary condition and
using it in equation (26), one easily sees that magnetic structures would take
a time $\Delta t \simeq 7$ yr to cross the distance from the base to the top of
the convective envelope.

During the above-estimated crossing time, diffusive phenomena can operate on
the material inside the magnetized zones, dispersing it into the environment; hence the frozen field hypothesis would not hold anymore. In
a simple diffusive treatment, a point-like density peak would be smeared out
over a length $\Delta x$ in a time $\Delta t$ such that:
$$
\Delta x^2 = D \Delta t
$$
where $D$ is some diffusion coefficient. According to \citet{pas} the effects
of magnetic and thermal diffusion and of viscosity are on average similar in the innermost
convective regions (so that there the magnetic Prandtl number
 becomes $P = \eta/\nu_m \simeq 1$). The dependency of the kinematic viscosity $\eta = \mu/\rho$ on temperature and density in a convective layer can be taken from \citet{vish}:
$$
\eta \simeq 2.2 \cdot 10^{-15}{T^{5/2}/\rho}, \eqno(27)
$$
hence we can assume, as a rough estimate, that $D \simeq 3 \nu_m \simeq 6.6 \cdot 10^{-15}({T^{5/2}/\rho})$. At the base of the
envelope (Table 1) this gives $D \simeq 1.85 \times 10^5  cm^2/sec$. For this
diffusion coefficient, in an envelope crossing time (7yr) the matter in the
tubes would be dispersed transversally to the tubes themselves over a distance
$\Delta x \simeq$ 6.4 10$^6$ cm, i.e. 64 Km. At larger radii in the envelope this value
would increase, as $\eta$ increases.

We can compare the above result with the typical dimensions of a thin
filamentary flux tube. In the Sun, for example, apart from large magnetized
areas with dimensions of thousands of Km, where the magnetic flux is $\Phi
\simeq 10^{22} - 10^{23}$ Mx, there is a {\it carpet} of fibril-like individual
tubes where the flux is much smaller \citep[$3\cdot 10^{17}$ to $3\cdot
10^{19}$ Mx, see][for details]{pri}. For an average flux of $\Phi = 3\times
10^{18}$ Mx, if magnetic zones are weaved of such individual filaments also at
the envelope bottom of an AGB star, we would have there fibrils with a radius
of $a_{env} \simeq {[\Phi/(\pi B)]^{1/2}}$, i.e. 60 Km. Normally, these thin fibrils would be
associated in bundles, with dimensions of thousands of Km.  However, in a
buoyancy crossing time of the envelope, the above very rough estimate shows
that the material internal to each fiber can diffuse away by a distance of the
same order as the fiber radius itself, thus probably making the flux tubes
incapable of grouping and floating to the surface. Although the diffusive
arguments presented above are very simple (actually rather qualitative), they
already suggest as probable that the transported material be dispersed through
the convective zone.

Notice that, by relaxing the frozen field assumption, the radial velocity in the
envelope would drop by a large factor with respect to Figure \ref{vrglobal}, so
that the dissipation of the magnetic fields in the environment is actually
guaranteed.

More quantitative studies of the erosion of magnetic
fields in turbulent media \citep{pm97} found that the tubes can become subject to
significant flux loss and to the creation of external current sheets
(transporting matter) over short timescales (about 1 month for the solar
convective layers). Moreover, it is  known \citep{mi92,rs01} that in a convective
environment the material transported by the tubes born in an underlying radiative
layer has a higher entropy than the non-magnetized envelope and that this entropy
contrast increases radially, as the entropy of the convective regions decreases; in
solar conditions, for fields below about $10^5$ G, this cannot be supported up to the
surface, so that even tubes surviving diffusion at the envelope base, would
later ``explode''. As an example, in the solar convective layers a tube with a
field like the one deduced above (2.65$\cdot 10^4$ G) would explode some 60$-$70
Million Km below the photosphere \citep{mi92,fan06}.

We want to note finally that the situation for tube disruption and matter diffusion
is very different in the underlying radiative regions, so that our initial hypothesis
of  almost {\it frozen} flux tubes is reasonable there. In order to show this, let us consider the
buoyancy velocity immediately below $r = r_{env}$. As we had no dissipative effects in the radiative layers on the radial velocity, it will probably be close to the Alfv\'{e}n speed estimated above (3.3 Km/sec); anyhow, from the estimate made before at the base of the envelope, we know it must be larger than 100 m/sec. Let's use this cautional value to be sure, thus obtaing an {\it upper limit} to the crossing time of the radiative layers. We can use formula (16), deriving $v_{r,P}$ from this last. It is then found that this crossing time is $\Delta t_{rad} \lesssim 1.25\times 10^8$ sec (about 4 years). Then, assuming magnetic flux conservation, we can estimate the tube dimension at $r_P$ from posing
$(a_{env}/a_{P})^2 = (r_{env}/r_P)^2$. This  yields $(a_{env}/a_{P})^2 \simeq $ 694, i.e. fibrils of $a_P$ = 8.65$\times$10$^3$ cm.
In order for the material to diffuse by such a distance in the estimated crossing time we would need a diffusion coefficient
$D = \eta \gtrapprox (8.65\times 10^3)^2/1.25\times10^8 \gtrapprox 0.6$ cm$^2$/sec. However, from the value of $\mu$ derived in Section 2 ($\mu \simeq$ 0.01) we know that $\eta$ is instead very small ($\eta = \mu/\rho \simeq 0.01/\rho_P$, i.e. 2.5$\times$ 10$^{-3}$ cm$^2$/sec). Hence, even assuming the lowest possible estimate for $v_{r,env}$, diffusion processes have no time to occur and this shows that our quasi-ideal MHD hypothesis, with almost frozen fields, is realistic.

Summing up, here we performed an analytical exact treatment of the MHD
equations under simple, two-dimensional, but rather realistic hypotheses
on the field geometry and on the matter density distribution in the layers external
to a H-burning shell in an AGB star. We showed that the buoyancy of magnetized 
structures provides an exact solution to the problem; this is so
both for the radiative and for the overlying convective zones.
In the radiative region pure advection (on time scales much faster than any diffusion
process) can yield the rapid crossing of the layer up to the convective envelope base,
thanks to the fact that the radial velocity grows as the second power of the radius.
As the physics of the expansion process is controlled by what occurs in these 
radiative zones above the H-burning shell, there we verified our 2D treatment with 
a dedicated 3D analysis of the velocity components, showing that the dependency on 
the second power of the radius remains valid, thus making our result quite robust.

For the envelope, unfortunately, a purely analytical and exact solution is not
possible, as it cannot include the effects of macroscopic eddy
motions characterizing turbulent convection. It would in any case result that
the motion is not compatible with pure advection and occurs on long time scales,
so that other phenomena, like magnetic, thermal and kinematic diffusion, or
flux tube explosion, would prevail over buoyancy, thus dispersing locally
any material carried by magnetic structures.

The radial (buoyant-advective) transport in the radiative layers 
is much faster than any diffusive phenomenon, but on average much slower than for pure
convection. These conditions were actually ascertained to be required by previous
studies on deep mixing in stars \citep{palm1,palm2,dm11}.

All the above characteristics converge in suggesting that magnetic buoyancy is
probably the most promising mixing mechanism so far devised for solving the
observational problems posed by the chemical and isotopic peculiarities of evolved
red giants.

The rare phenomena of Li production, occasionally observed in RGB stars and so far
unexplained, might be another puzzle for which rapidly emerging magnetic structures
might offer a solution. The fast transport guaranteed by magnetic buoyancy might
indeed mimic the so-called Cameron-Fowler mechanism \citep[see e.g.][and references 
therein]{palm2}, saving to the envelope $^7$Be, the unstable nucleus that is the 
parent of Li, before
it can disappear through strong and weak interactions in the radiative zones.
Once in the envelope, it would then decay to Li, guaranteeing its enrichment.
This is however a subject that requires further scrutiny, after the $e$-capture
rate of $^7$Be has been recently remarkably revised \citep{sim}.

{\bf Acknowledgements}. The rather long development of this work was permitted by INFN, through the financial support of its Group 3 (Nuclear Physics and Astrophysics) and through the ERNA collaboration. We acknowledge many stimulating discussions with G. J. Wasserburg and S. N. Shore, as well as useful suggestions from a competent referee.

\onecolumn
\renewcommand{\theequation}{A-\arabic{equation}}
\setcounter{equation}{0}  
\section*{APPENDIX}  
The general conditions discussed in the text on the components of the magnetic
field and velocity are the only approximations we need. With those
assumptions, the MHD equations [i.e. equations $(1) - (4)$ of Section 2], in the plane polar coordinates $r$ and $\varphi$, are:
\begin{eqnarray}
\frac{\partial \varrho}{\partial t}+\frac{\partial (\varrho v_r)}{\partial r}
+\frac{1}{r}\frac{\partial (\varrho v_{\varphi})}{\partial \varphi}+\frac{\varrho
v_r }{r}=0\label{cont}\\[0.3cm]
\varrho\left(\frac{\partial v_r}{\partial t}+v_r\frac{\partial v_r}{\partial
r}+\frac{v_{\varphi}}{r}\frac{\partial v_{\varphi}}{\partial
\varphi}-\frac{v_{\varphi}^2}{r}  + \frac{\partial \Psi}{\partial r} - c_d v_r \right) + \frac{\partial P}{\partial r} \nonumber\\
-\mu\left(\frac{\partial^2 v_r}{\partial r^2} + {\frac{1}{r^2}} \frac{\partial^2
v_{\varphi}}{\partial {\varphi}^2} + {\frac{1}{r}} \frac{\partial v_r}{\partial
r}-\frac{2}{r^2} \frac{\partial v_{\varphi}}{\partial \varphi}
-\frac{v_r}{r^2}\right)
+\frac{1}{4\pi}\frac{B_{\varphi}}{r}\left(\frac{\partial (r
B_{\varphi})}{\partial r}-\frac{\partial B_r}{\partial
\varphi}\right)=0\label{NSr}\\[0.3cm]
\varrho\left(\frac{\partial v_{\varphi}}{\partial t}+v_r\frac{\partial
v_{\varphi}}{\partial r}+\frac{v_{\varphi}}{r}\frac{\partial v_{\varphi}}{\partial
{\varphi}}+\frac{v_r v_{\varphi}}{r} + \frac{1}{r}\frac{\partial \Psi}{\partial {\varphi}} - c_d v_{\varphi} \right) + \frac{1}{r}\frac{\partial P}{\partial
{\varphi}}
\nonumber\\
 -\mu \left(\frac{\partial^2 v_{\varphi}}{\partial
r^2}+\frac{1}{r^2}\frac{\partial^2 v_{\varphi}}{\partial
\varphi^2}+\frac{1}{r}\frac{\partial v_{\varphi}}{\partial
r}+\frac{2}{r^2}\frac{\partial v_r}{\partial \varphi}
-\frac{v_{\varphi}}{r^2}\right)
-\frac{1}{4\pi}\frac{B_r}{r}\left(\frac{\partial (r B_{\varphi})}{\partial
r}-\frac{\partial B_r}{\partial
\varphi}\right)=0\label{NSth}\\[0.3cm]
\frac{\partial B_r}{\partial t}-\frac{1}{r}\frac{\partial (v_r
B_{\varphi}-v_{\varphi}B_r)}{\partial \varphi} -\nu_m\left(\frac{\partial^2
B_r}{\partial r^2}+\frac{1}{r^2}\frac{\partial^2 B_r}{\partial \varphi^2}
+\frac{1}{r}\frac{\partial B_r}{\partial r}-\frac{2}{r^2}\frac{\partial
B_{\varphi}}{\partial \varphi}-\frac{B_r}{r^2}\right)=0\label{eqMr}\\[0.3cm]
\frac{\partial B_{\varphi}}{\partial t}+\frac{\partial (v_r
B_{\varphi}-v_{\varphi}B_r)}{\partial r}-\nu_m\left(\frac{\partial^2
B_{\varphi}}{\partial r^2}+\frac{1}{r^2}\frac{\partial^2 B_{\varphi}}{\partial
\varphi^2} +\frac{1}{r}\frac{\partial B_{\varphi}}{\partial
r}+\frac{2}{r^2}\frac{\partial
B_r}{\partial \varphi}-\frac{B_{\varphi}}{r^2}\right)=0\label{eqMth}\\[0.3cm]
\frac{\partial B_r}{\partial r}+\frac{1}{r}\frac{\partial B_{\varphi}}{\partial
\varphi}+\frac{B_r}{r}=0\label{eq8}
\end{eqnarray}
Here the definitions for the symbols used can be found in the main text.
In equations (\ref{NSr}) and  (\ref{NSth}) one has $\displaystyle\frac{\partial
P}{\partial \varphi}=0$, $\displaystyle\frac{\partial
\Psi}{\partial \varphi}=0$, $\displaystyle\frac{\partial
P}{\partial r}= \frac{{\rm d} P}{{\rm d} r}$ and  $\displaystyle\frac{\partial
\Psi}{\partial r}= \frac{{\rm d} \Psi}{{\rm d} r}$.\\[0.3cm]
Equation (\ref{eq8}) yields $\displaystyle\frac{\partial
B_{\varphi}}{\partial \varphi}=0$ since we assumed $B_r=0$.\\[0.3cm]
Equation (\ref{eqMr}) yields $\displaystyle\frac{\partial v_r}{\partial
\varphi}=0$ since we assumed $\nu_m=0$.\\ [0.3cm]
Equation (\ref{cont}) becomes a first-order ordinary differential equation in $v_r$ as a function of $r$, since we assumed:
 $\displaystyle\frac{\partial v_{\varphi}}{\partial \varphi}=0$ and $\varrho= C \cdot r^k$;
in the conditions discussed in the text, the constant $C$ is $C$ = $\varrho_P/r_P^k$.
From it, we obtain that: $$v_r=\displaystyle\frac{{\rm d}w(t)}{{\rm
d}t}\, {r}^{-(k+1)}$$
where $w$ is a function whose dimensions are
[L$^{k+2}$]. Notice that this solution for the radial velocity is independent of drag,
as it is obtained without reference to equation (\ref{NSr}).
Now equation (\ref{eqMth}) becomes a linear, first-order, partial
differential equation in $B_{\varphi}$, whose general solution
can be written as:
$$
B_{\varphi}=\Phi(\xi){r}^{k+1},\quad \left[\xi=-(k+2)w(t)+{r}^{k+2}\right].
$$
where $\Phi(\xi)$  is a function of $\xi$, measured in units (Gauss $\cdot$ cm$^{-(k+1)}$].
As $\displaystyle\frac{\partial P}{\partial \varphi}=0$,
then equation (\ref{NSth}) reduces to a non-constant-coefficient homogeneous
linear diffusion equation of the second order in the unknown $v_{\varphi}(t,r)$.
\\ [0.3cm]

\end{document}